\documentclass[3p,12pt]{elsarticle}
\usepackage{hyperref}

\usepackage{amssymb,amsmath}
\usepackage{graphicx}  
\usepackage{multirow}
\usepackage{graphicx}
\usepackage{caption}
\usepackage{subcaption}
\usepackage{algorithm}
\usepackage{algpseudocode}
\newtheorem{definition}{Definition}
\DeclareMathOperator*{\argmax}{arg\,max}

\begin{document}
\begin{frontmatter}
\title{Implicit Contextual Integrity in Online Social Networks}
\author[a1]{Natalia Criado}
\author[a2]{Jose M. Such}
\address[a1]{School of Computer Science\\
Liverpool John Moores University, Liverpool, UK\\
n.criado@ljmu.ac.uk}
\address[a2]{School of Computing and Communications\\
Infolab21, Lancaster University, Lancaster, UK\\
j.such@lancaster.ac.uk}

\begin{abstract}
Many real incidents demonstrate that users of Online Social Networks need mechanisms that help them manage their interactions by increasing the awareness of the different contexts that coexist in Online Social Networks and preventing them from exchanging inappropriate information in those contexts or disseminating sensitive information from some contexts to others. Contextual integrity is a privacy theory that conceptualises the appropriateness of information sharing based on the contexts in which this information is to be shared. Computational models of Contextual Integrity assume the existence of well-defined contexts, in which individuals enact pre-defined roles and information sharing is governed by an explicit set of norms. However, contexts in Online Social Networks are known to be implicit, unknown a priori and ever changing; users relationships are constantly evolving; and the information sharing norms are implicit. This makes current Contextual Integrity models not suitable for Online Social Networks.

In this paper, we propose the first computational model of \emph{Implicit} Contextual Integrity, presenting an information model for Implicit Contextual Integrity as well as a so-called Information Assistant Agent that uses the information model to learn implicit contexts, relationships and the information sharing norms in order to help users avoid inappropriate information exchanges and undesired information disseminations. Through an experimental evaluation, we validate the properties of the model proposed. In particular, Information Assistant Agents are shown to: (i) infer the information sharing norms even if a small proportion of the users follow the norms and in presence of malicious users; (ii) help reduce the exchange of inappropriate information and the dissemination of sensitive information with only a partial view of the system and the information received and sent by their users; and (iii) minimise the burden to the users in terms of raising unnecessary alerts.

\end{abstract}


\end{frontmatter}


\section{Introduction}
Online Social Networks (OSNs) have been a source of privacy concerns and issues since their early days \cite{gross2005information}. These privacy concerns have even increased along the past decade due to many real privacy incidents being echoed in the media and users being more aware of potential privacy issues \cite{hargittai2010facebook,stutzman2013silent}. Yet there is a lack of effective privacy controls that allow users to satisfactorily manage their privacy in OSNs \cite{strater2008strategies}
. In particular, the exchange of inappropriate information and the undesired dissemination of sensitive information across OSNs are very common and represent one of the major concerns for users. These inappropriate exchanges and undesired disseminations have not only caused serious privacy incidents --- e.g., users have lost their jobs \cite{pike2011fired}, have been outed and faced threats to sever family ties \cite{fowler}, have ended their marriages \cite{stevens}, etc. --- but also facilitated other activities such as social phishing \cite{jagatic2007social}, identity theft \cite{bilge2009all}, cyberstalking \cite{lyndon2011college}, and cyberbullying \cite{ruedy2007repercussions}.


Some voices argue that this is mainly due to the fact that users are no longer able to share information differently for different contexts or spheres of life (friends, work, etc.) in the cyber world, as they would usually do in the physical world \cite{wang2011regretted}. There are many examples in which this is of crucial importance: photos that depict users in embarrassing situations, indecorous comments, events that reveal some political affiliations, etc. In all these examples, the specific context determines whether or not the exchange of information is appropriate --- e.g., one may be willing to share her political affiliations with friends but not with workmates.

Contextual integrity \cite{nissenbaum2004privacy} is a modern privacy theory that conceptualises the appropriateness of information sharing based on the contexts in which this information is to be shared. In particular, contexts are defined considering a set of individuals playing particular roles and a number of norms that govern information sharing among them. 
Contextual integrity is said to be maintained --- meaning that there are no privacy breaches --- whenever these information sharing norms are upheld. The information sharing norms have two main purposes: (i) determine what information is appropriate to mention in a particular context, and (ii) dictate what information can be transmitted from one party to another or others according to the roles enacted by these parties within and across different contexts. 

Computational models of contextual integrity have been recently proposed in the related literature \cite{barth2006privacy,krupa2012handling}. Following contextual integrity theory, they assume the existence of well-defined contexts, in which individuals enact pre-defined roles and information sharing is governed by an explicit set of norms. However, contexts in OSNs are ``implicit, ever changing and not a priori-known'' \cite{danezis2009inferring}. In particular, the information sharing norms are known to be \emph{implicit} in OSNs \cite{raynes2006hyperfriends,vorvoreanu2009perceptions}, i.e., they define the behaviour that is consistent with the most common behaviour. Moreover, roles are dynamic and may not be known a priori --- i.e., relationships among individuals in OSNs are constantly evolving \cite{burke2014growing}. All of these reasons make explicit contextual integrity and the computational models based on it not suitable for OSNs.



In this paper, we present the first computational model of \emph{implicit} Contextual Integrity for OSNs, which includes an information model and an agent model for a kind of agents we call \emph{Information Assistant Agents} (IAAs). IAAs are capable of learning contexts and their associated information sharing norms even if these are implicit or unknown a priori. 
Each IAA monitors the information exchanges of its user and based on this it infers: (i) the different contexts in which information sharing is to happen; (ii) the relationships among the individuals in each context; and (iii) the information sharing norms of each context. If IAAs detect a potential violation of the information sharing norms, they alert their users, who have the last word on whether sharing the information or not. Through an experimental evaluation we demonstrate that IAAs are able to significantly mitigate the exchange of inappropriate information and the undesired dissemination of sensitive information while minimising the number of unnecessary alerts raised.

This paper is organised as follows. Section \ref{sec:problem} states the problem we tackle in this paper detailing the exchange of inappropriate information and the undesired dissemination of sensitive information in OSNs. Section \ref{sec:information} proposes the information model that is used by IAAs to promote contextual integrity by alerting users when their actions may lead to the violation of the information sharing norms. The IAA model and its modules are described in Section \ref{sec:agent}. Section \ref{sec:examples} illustrates the performance of IAAs in a set of examples. Section \ref{sec:evaluation} details the experiments we conducted to evaluate our proposal and the results we obtained. Section \ref{sec:discussion} discusses related work. Finally, Section \ref{sec:conclusion} includes some concluding remarks as well as future work.

\section{Problem Statement}\label{sec:problem}



IAAs aim to avoid two main privacy threats in OSNs: the exchange of inappropriate information, and the undesired dissemination of previously exchanged information. Specifically, they alert their users that sharing some information may lead to some of these threats. 
Next, we describe these threats in more detail.
\subsection{Inappropriate Information Exchanges}
Each context has its own appropriateness norms that determine which information can be mentioned inside each context. For example, one may not mention her political views in a work context, but she may do so in a family context \cite{wang2011regretted}. IAAs use the information that their users exchange with other users in one context to infer the appropriateness norms of this context. Specifically, the information that is frequently mentioned by the members of a context is considered as appropriate whereas information that is never or rarely mentioned is considered as inappropriate. For instance, if most people do not mention their political views at work, it could be inferred this is not an appropriate topic to exchange in a work context.

Besides the appropriateness norms of each context, there are situations in which people decide to exchange information that may be seen as inappropriate. One of the main reasons that explain this is the creation and reciprocation of close relationships \cite{greene2006self}. Indeed, there are empirical studies that demonstrate the fact that reciprocated communication is the dominant form of interaction in OSNs \cite{DBLP:conf/socialcom/ChengRMK11}. Accordingly, IAAs take into account the appropriateness of the information that has been exchanged with each user to determine when inappropriate information is being exchanged to reciprocate a close friend or to create a close relationship. 

\subsection{Undesired Information Disseminations}
Disseminations occur when information disclosed in one context travels to another context. That is, disseminations are inter-context disclosures while exchanges (as stated above) are intra-context disclosures. Obviously, if the information to be disclosed is already known in the contexts were it may be disclosed, then the disclosure of this information in these contexts cannot entail any new privacy risk. However, disseminations may potentially be undesired and hazardous when they entail the disclosure of sensitive information that was previously unknown in a context \cite{strahilevitz2005social}. Indeed, studies on regrets associated to users' posts on OSN highlight the fact that revealing secrets of others is one of the main sources of regret \cite{wang2011regretted}. For instance, there was a recent case in which the sexuality of a person was leaked from her friends context to her family context where her sexuality was previously unknown, causing her being outed and facing threats to sever family ties \cite{fowler}.



A first-line defence against undesired disseminations may be avoiding sharing sensitive information in contexts in which there are people that could disseminate the information to other contexts in which this information is previously unknown. Whether these people decide to disseminate the information or not may depend on the relationship they have to others. That is, people usually have confidence relationships with others with whom they decide to share sensitive information expecting them to keep it secret. One can share some of her deepest secrets with her husband but this may not mean her husband would disseminate this information to other contexts. Thus, IAAs take into account the knowledge of the information that has been exchanged with each user to determine when sensitive information is being exchanged to reciprocate a trusted friend or to create/maintain trust relationships.

\section{Information Model for Implicit Contextual Integrity}\label{sec:information}
In this section, we describe the information model for implicit contextual integrity, which is used by an IAA (further detailed in Section \ref{sec:agent}) to assist its user, denoted by $\alpha$, in sharing information in the OSN. 

The information model considers 
a set of users $\mathcal{U}=\{\alpha,u_{1},...,u_{n}\}$  who are members of and communicate in an OSN. Information exchanged among users is formalised as messages denoted by tuples $\langle g,R,T,S\rangle$; where $g \in \mathcal{U}$ is the sender user, $R\subseteq \mathcal{U}$ is the set formed by the receiver users, and $T$ and $S$ represent the information that is sent.  In general, messages represent any exchange of information in an OSN, where this information may be photos, comments, private messages, and so on. Let $\mathcal{T}=\{t_{1},...,t_{m}\}$ be a set of key topics or categories that can be discussed in messages. Then, $T\subseteq\mathcal{T}$ is the set of topics discussed in a message, and $S\subseteq\mathcal{U}$ are the users \textit{tagged} (i.e., associated) in this message. Informally, we define a context in our problem as a group of users who interact through a specific OSN. Formally, a context $C$ is a set of users ($C\subseteq \mathcal{U}$). We denote by $\mathcal{C}=\{C_{1},\ldots,C_{l}\}$ the set of contexts in which user $\alpha$ interacts with other users. We denote by $\mathcal{C}_{\beta}$ the set of contexts that $\alpha$ and $\beta$ share:
\[\mathcal{C}_{\beta}=\{C_{i}|\forall C_{i}\in\mathcal{C}:\beta\in C_{i}\}\]

\subsection{Appropriateness}
As aforementioned, we propose that IAAs infer the information sharing norms from the information that is exchanged in the OSN. This section includes the main definitions used by IAAs when inferring the appropriateness norms. In particular, we start by defining the individual appropriateness values; i.e., users' judgement of what is appropriate or inappropriate. Then, we define context appropriateness values; i.e., contexts' appropriateness judgement. Finally, we define the appropriateness of messages. 

\begin{definition}[Individual Appropriateness]
Given a user $\beta\in\mathcal{U}$ and a topic $t\in \mathcal{T}$, we define the \textit{individual appropriateness} $a_{\beta}^{t}$ as a real value within the $[0,1]$ representing the likelihood that user $\beta$ approves to exchange information about topic $t$. 
\end{definition}

These likelihoods are estimated as frequentist probabilities according to the information contained in messages that $\alpha$ receives from other users. If an IAA is created to support a user that is new to the OSN, then there is not prior knowledge that the IAAs can use to infer the appropriateness of topics. In absence of prior knowledge, the IAA cannot determine if exchanging  information about one topic is appropriate or not and, as a consequence, $\alpha$ initialises the likelihood values to 0.5, which means that it is equally probable that something is appropriate or inappropriate \cite{ross2002fuzzy,iba2009applied,dasgupta1997evolutionary}. If an IAA is created to support a user that has been a member of a OSN for a while, then the individual appropriateness values can be initialised based on the messages sent by each user $\beta$ (i.e., messages sent by $\beta$ to $\alpha$ can be used to infer to what extent $\beta$ approves to exchange information about the different topics).

Later, the IAA updates these probabilities as it receives messages. For example, if Alice (denoted by user $\alpha$) receives a message from Bob (denoted by user $\beta$) in which a topic such as a dirty joke (denoted by $t$) is mentioned, then this may suggest that Bob considers that making dirty jokes is appropriate and, therefore, the likelihood $a_{\beta}^{t}$ is increased accordingly.

\begin{definition}[Appropriateness Increase]
Given an individual appropriateness value $a_{\beta}^{t}$, we define the appropriateness increase as follows:
\[
 f_{AppropriatenessIncrease}(a_{\beta}^{t})=min\{1,a_{\beta}^{t}+\bigtriangleup(a_{\beta}^{t})\}
\]
where $\bigtriangleup:[0,1]\rightarrow[0,1]$ is an update function. 
\end{definition}
Information sharing norms of a context may change as time goes by; e.g., revealing personal information about couples may be appropriate for groups of young friends but it may become inappropriate when these friends grow up and they are engaged in stable relationships. In absence of information that sustains the approval of exchanging information about some topic, it seems reasonable to consider that it is becoming less usual and finally even disapproved. Therefore, the likelihoods that represent appropriateness also decay with time.
\begin{definition}[Appropriateness Decay]
Given an individual appropriateness value $a_{\beta}^{t}$, we define the appropriateness decay as follows:
\[
f_{AppropriatenessDecay}(a_{\beta}^{t})=
max\{0, a_{\beta}^{t}-\bigtriangledown(a_{\beta}^{t})\}
\]
where $\bigtriangledown:[0,1]\rightarrow[0,1]$ is an update function.
\end{definition}
If $\beta$ exchanges information about one topic, it means that it approves to exchange information about this topic. However, if $\beta$ does not exchange information about this topic, it does not necessarily mean $\beta$ disapproves to exchange information about this topic. As a consequence, we define that for any likelihood value $\delta\in[0,1]$, $\bigtriangleup(\delta)\gg \bigtriangledown(\delta)$.

Given a context $C\in \mathcal{C}$ and a set of topics $T\subseteq \mathcal{T}$, we denote by $A_{C}^{T}$ the set of the likelihoods that users in $C$ approve to exchange information about topics in $T$:
\[A_{C}^{T}=\{a_{\beta}^{t}|\beta\in C\wedge t\in T\}\]

In the information model, the likelihoods that other users approve to exchange information about topics are used to determine when a communicative action (i.e., a message sent by its user) can be seen as inappropriate. Assume that $\alpha$ wants to send a message ($\langle \alpha,R,T,S\rangle$) in which it exchanges information about a set of topics $T$. Assume also that the message belongs to just one context $C\in \mathcal{C}$. Thus, $\alpha$ uses its likelihoods that other users in context $C$ approve to exchange information about topics in $T$ to determine the appropriateness of this message. For example, Alice wants to send a message to Bob in which she makes a dirty joke. Given that Bob is a workmate of Alice (the work context is denoted by $C_{Work}$), the IAA uses the likelihoods that other workmates approve to exchange dirty jokes to determine the appropriateness of this message. 
\begin{definition}[Context Appropriateness]
Given a context $C\in \mathcal{C}$ and a set of topics $T\subseteq\mathcal{T}$, we define the context appropriateness as follows:
\[f_{ContextAppropriateness}(C,T)= f_{Combine}(A_{C}^{T})
\]
where $f_{Combine}:{[0,1]^{n}}\rightarrow[0,1]$ is a function that combines the different likelihood values into a single value. 
\end{definition}

$f_{Combine}$ may be defined differently according to the requirements of each user. For example, extremely cautious users may be interested in defining it as the minimum to avoid mentioning any topic that may be seen as inappropriate by someone.

Let us assume now that $\alpha$ wants to send a message but that the context of this message is unknown a priory. We determine the context of this message considering both the set of receiver users and the set of users associated with the information. Informally, we define the context of a message as the context in which the overlap among its members and the users involved in the message (i.e., the receiver and the tagged users) is maximal. For example, Alice wants to send a message (denoted by $\langle \alpha,R,T,S\rangle$) to Mary (denoted by user $\gamma$ so that $R=\{\gamma\}$) in which she mentions that she, Bob and Charlie (denoted by user $\pi$), who is also a  workmate, will attend a training event (thus $S=\{\alpha, \beta, \pi\}$). Mary is both Alice's best friend and a workmate; thus, this message might belong to the work and friend contexts. Given that the majority of the people involved in the message belong in the work context, it is highly possible that this message belongs in the work context.
\begin{definition}[Message Context]
Given a message $\langle \alpha,R,T,S\rangle$, we define the contexts of a message as follows:
\[f_{Context}(\langle \alpha,R,T,S\rangle)=\argmax_{C\in\mathcal{C}}|C \cap (R \cup S)|
\]
\label{def:context}
\end{definition}
In our example, Alice wants to send a message to Mary in which she mentions that she, Bob and Charlie will attend a training event. In this case, the set of contexts in which Alice interacts with other persons are: the work context denoted by $C_{Work}$ (note that $\beta, \pi, \gamma\in C_{Work}$); and the friend context denoted by $C_{Friend}$ (note that $\gamma\in C_{Friend}$ and $\beta, \pi\not\in C_{Friend}$). According to definition \ref{def:context}, the context of this message is $C_{Work}$ since $|C_{Work} \cap \{\beta, \pi,\alpha,\gamma\}|=3$ is greater than $|C_{Friend} \cap \{\beta, \pi,\alpha,\gamma\}|=1$.

Note, however, that the previous function may also return more than one context in other situations. For example, Alice sends a message to Mary who is a workmate and a friend (remember $\gamma\in C_{Friend}$ and $\gamma\in C_{Work}$) and this message may belong to the work and the friend contexts. Thus, any function that calculates the appropriateness of messages must take into account that a message may belong to more than one context and that this message can be interpreted differently in each context. IAAs want to prevent their users from exchanging information about topics that could be seen as inappropriate in any of the possible contexts. Thus, they calculate the appropriateness of messages considering the context in which this will be less appropriate. 
\begin{definition}[Message Appropriateness]
Given a message $\langle \alpha,R,T,S\rangle$, we define the message appropriateness as follows:

\[
f_{Appropriateness}(\langle \alpha,R,T,S\rangle)=
\min_{C\in f_{Context(\langle \alpha,R,T,S\rangle)}} f_{ContextAppropriateness}(C,T)
\]

\end{definition}

As previously mentioned, people might be willing to exchange inappropriate information with those ones that have previously exchanged inappropriate information with them. For example, Alice may be willing to exchange a dirty joke with Bob (remember that Bob is a work mate), even if this may be seen as inappropriate in the work context because Bob and Alice have a close relationship and it is common that they exchange this kind of jokes. To take this into account, the IAA needs to know which information has been exchanged with the rest of users. We define the appropriateness exchange list with $\beta$ (denoted by $\mathcal{A}_{\beta}$) as the set formed by the appropriateness of all messages that have been exchanged between $\alpha$ and $\beta$. Note that this list contains the appropriateness of messages (not the messages) and, as consequence,  the IAA does not store messages sent or received by its user. In absence of prior knowledge, this list is initialised as the empty list. This appropriateness exchange list is used by the IAA to know with whom inappropriate information has been exchanged in the past and, on the basis of that information, to determine which users have a close relationship with its user.

\subsection{Information Dissemination}
Disseminations occur when the information that has been exchanged in one context is leaked in contexts where this information is unknown. Thus, it is crucial to enable IAAs with capabilities that allow them to anticipate dissemination before it happens. We propose that IAAs use the messages sent and received in which users are tagged explicitly (e.g., a user tagged in a photo) to calculate the knowledge of the associations between users and topics (e.g., the relationship between Bob and the topic homosexuality) and determining which ones are unknown. Again, we start by defining the individual knowledge values, then we define context knowledge values, and finally we define the knowledge of messages.

\begin{definition}[Individual Knowledge]
Given any two different users $\gamma,\beta\in\mathcal{U}$ and a topic $t\in \mathcal{T}$, we define the individual knowledge $k_{\beta}^{\gamma,t}$ as a real value within the $[0,1]$ interval representing the likelihood that $\beta$ knows the association of user $\gamma$ with topic $t$.
\end{definition}
As in case of the individual appropriateness values, the individual knowledge values can be initialised considering the existence of previous messages received by $\alpha$. In the absence of this information, the individual knowledge values are initialised to 0.5.

As $\alpha$ receives messages from other users, the IAA uses the information contained in messages for updating these likelihoods. For example, if Alice receives a message ($\langle \beta,R,T,S\rangle$) in which Bob ($\beta$) tells to Charlie and Alice ($R=\{\alpha,\pi\}$) that Mary ($S=\{\gamma\}$) has been promoted ($T=\{t\}$; where $t$ denotes the promotion topic), then the IAA has some evidence that sustains the fact that Bob knows the association of Mary with the promotion topic. Moreover, the IAA also infers that the rest of receiver users (Charlie) know the association of Mary with the promotion topic. However, it may be the case that Bob forwarded this message without paying much attention to it or analysing it properly. Similarly, it may be the case that Charlie never reads this message. Because of this, the IAA cannot be completely sure that users on $S$ and $R$ know the association of $\gamma$ with topic $t$ and it cannot set knowledge likelihoods to 1. Thus, the IAA can only increase the \emph{likelihood} that users on $S$ and $R$ know the association of $\gamma$ with topic $t$.
\begin{definition}[Knowledge Increase]
Given an individual knowledge value $k_{\beta}^{\gamma,t}$, we define the knowledge increase as follows:

\[f_{KnowledgeIncrease}(k_{\beta}^{\gamma,t})=min\{1,k_{\beta}^{\gamma,t}+\bigtriangleup(k_{\beta}^{\gamma,t})\}
\]
\end{definition}
The information that we know may change as time goes by (e.g., the relationship status of our Facebook friends changes along time) or, even, users may forget information (e.g., we can forget about the hundreds of Facebook events to which we have been invited), thus the knowledge likelihoods decrease lightly with time. 
\begin{definition}[Knowledge Decay]
Given an individual knowledge value $k_{\beta}^{\gamma,t}$, we define the knowledge decay as follows:
\[
f_{KnowledgeDecay}(k_{\beta}^{\gamma,t})=max\{0, k_{\beta}^{\gamma,t}-\bigtriangledown(k_{\beta}^{\gamma,t})\}
\]
\end{definition}

Given a context $C\in\mathcal{C}$, user $\gamma\in\mathcal{U}$ and a set of topics $T\subseteq\mathcal{T}$ we denote by $K_{C}^{\gamma,T}$ the set of likelihood that the association of user $\gamma$ with topics in $T$ is known by users in $C$:
\[K_{C}^{\gamma,T}=\{k_{\beta}^{\gamma,t}|\beta\in C\wedge t\in T\}\]

Assume that $\alpha$ wants to send a message ($\langle \alpha,R,T,S\rangle$) to $\beta$ ($R=\{\beta\}$) in which user $\gamma$ ($S=\{\gamma\}$) is associated with topics $T$. Upon receiving the message $\beta$ would know that association and could disclose it in other contexts. This may cause that $\beta$ disseminates the information received from $\alpha$. This dissemination can be critical if $\beta$ reveals the association of $\gamma$ with topics in $T$ in contexts where $\gamma$ belongs and this association is unknown. Assume that $\gamma$ and $\beta$ belong to a context $C$, then $\alpha$ can calculate the knowledge within context $C$ of the relationship between $\gamma$ and topics in $T$ by considering the likelihood that the association of user $\gamma$ with topics $T$ is unknown by users in $C$. For example, Alice knows that Mary is pregnant. No one else has talked about Mary's pregnancy at work. Thus, Alice assumes that this information about Mary is unknown at work.
\begin{definition}[Context Knowledge]
Given a context $C\in \mathcal{C}$, we define the knowledge within context $C$ of the relationship between $\gamma\in\mathcal{U}$ and topics in $T\subseteq\mathcal{T}$ as follows:
\[f_{ContextKnowledge}(C,\gamma,T)= f_{Combine}(K_{C}^{\gamma,T}) \]
\end{definition}

Given that any message can be sent to a set of receiver users $R$ and may tag a set of users $S$, the previous function should be extended to several receiver and tagged users. This also entails that there may be several contexts where information can be disseminated. The IAA wants to avoid disseminations of information that can be unknown in some context. Thus, the IAA calculates the knowledge considering the least known information that can be disseminated. For example, Alice may want to send a message to her family ($S$) in which she tells about Mary's pregnancy. Charlie is her brother-in-law ($\pi\in R$). Charlie also works at the same company (i.e., Mary and Charlie share the work context) and, as a consequence, Charlie might reveal the information about Mary's pregnancy at work. Thus, this message may entail the dissemination of unknown information at the work context.
\begin{definition}[Message Knowledge]
Given a message $\langle \alpha,R,T,S\rangle$, we define the knowledge of a message as follows:
\[
f_{Knowledge}(\langle \alpha,R,T,S\rangle)=\min_{\substack{
            \forall s\in S, \forall r\in R \\
            \forall C\in(\mathcal{C}_{s}\cap\mathcal{C}_{r})}} f_{ContextKnowledge}(C,s,T)\]
where $\mathcal{C}_{s}\cap\mathcal{C}_{r}$ represents the contexts shared by $s$ and $r$.
\end{definition}

As previously mentioned, the creation and maintenance of trust relationships explains the fact that people might be willing to exchange unknown information which those ones who have previously exchanged unknown information with them. The IAA need to know the knowledge of the information that has been exchanged with the rest of users. We define as the knowledge exchange list with $\beta$ (denoted by $\mathcal{K}_{\beta}$) as the set formed by the knowledge of all messages exchanged between $\alpha$ and $\beta$. In absence of prior knowledge, this list is initialised as the empty list. The knowledge exchange list is used by the IAA to determine which users have a trust relationship with its user\footnote{Note that having a close relationship is not the same as having a trust relationship; thus, we need to maintain two independent exchange lists.}. Exchanges of unknown information with these trusted friends would be justified by the disclosure reciprocity phenomenon. For example, the fact that Alice tells Charlie the information about Mary pregnancy (which may be disseminated in the work context) would be justified by the fact that Charlie is a trusted friend of Alice.

\section{Information Assistant Agent model}\label{sec:agent}
\begin{figure}
\center{
\includegraphics[trim = 0mm 225mm 0mm 0mm, clip,width=0.75\textwidth]{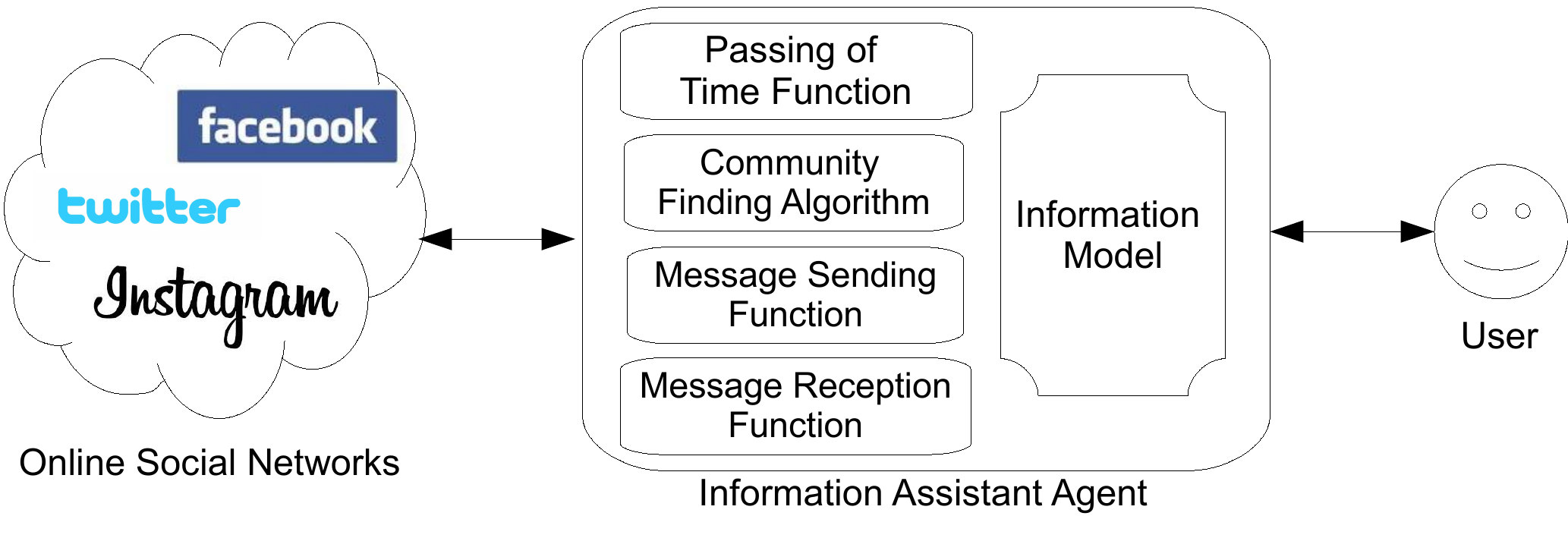}}
\vspace{-10pt}
\caption{Components and placement of an Information Assistant Agent.}
\label{fig:architecture}
\end{figure}

Figure \ref{fig:architecture} depicts the different modules that make up an IAA. We can see that it is aimed to be placed between one user and one OSN\footnote{Note that our model does not consider explicitly the existence of multiple interacting OSNs. However, multiple interacting OSNs can be combined using the approach proposed in \cite{buccafurri2014driving} and \cite{buccafurri2013supporting}.}. Thus, an IAA is responsible for managing the interactions (i.e., the messages sent and received) of a single user in an OSN. An IAA could be implemented on top of an OSN infrastructure (or Social Network Service) as a Social Networking App running in a trusted server or it could be implemented as a local Social App that acts as a proxy to access the OSN (e.g., a mobile Social App). Note that as stated in the previous section, an IAA does not store messages sent/received by their users, which helps to minimise storage space for the information model, which may be important to implement an IAA as a local App for resource-constrained devices like smartphones. Note also that IAAs utilise information that is currently available to users in mainstream OSNs and their applications ---i.e., the content posted and read by users (e.g., tweets) and the friendship relationships (e.g., the following and follow relationships). 

An IAA is composed of 5 different modules: (i) community finding algorithm; (ii) passing of time function; (iii) message sending function; (iv) message reception function; and (v) the information model. 
We already detailed in the previous section the information model, so this section focuses on explaining the rest of modules, how the information model interacts with them, and a description of an IAA's life-cycle.

\subsection{Community Finding Algorithm} In OSNs, the individuals that belong to a particular context are commonly referred to as a \textit{community} \cite{girvan2002community}. 
These \textit{communities} \cite{girvan2002community} are natural divisions of network users into densely connected subgroups. Many OSNs support some notion of community by means of allowing groups of users (e.g., Facebook groups or Google+ circles). There are even methods and ready-to-use tools that obtain them automatically using unsupervised machine learning algorithms \cite{rosvall2007information}. For example, IAAs could make use of existing community finding algorithms to extract the different communities, such as Infomap \cite{rosvall2007information}, which is known to perform accurately in OSNs like Facebook \cite{fogues2014bff} and is the algorithm IAAs use for the examples and evaluation we carried out, as described later on in the paper. Note that OSN infrastructures only allow each user to have a partial view of the OSN (that contains his/her friends and mutual friends). That is, each IAA needs to elicit the contexts in which its user is involved by considering a partial view of the network (also known as ego-network). 


The community finding algorithm can be executed by IAAs on the background without interfering with the users' actions. Specifically, IAAs can recalculate the communities at regular time intervals or after an event, e.g., any time there is a change in the friend network. 


\subsection{Passing of Time}
The pseudo code corresponding to this function is depicted in Algorithm \ref{alg:passing}. At this step, an IAA updates likelihoods as the times goes by. Specifically, the IAA applies the decay functions to the appropriateness likelihoods to model the fact that, in absence of new information that sustains the appropriateness of exchanging information about a topic, this appropriateness degrades over time (see line 3). Similarly, the IAA applies the knowledge decay function, which models the fact that, in absence of new information that reveals the association between a user and a topic, the knowledge of such association decreases (see line 6).
\begin{algorithm}
\begin{small}
\begin{algorithmic}[1]
\ForAll{$\beta\in\mathcal{U}$}
\ForAll{$t\in\mathcal{T}$}
\State $a_{\beta}^{t}\leftarrow  f_{AppropriatenessDecay}(a_{\beta}^{t})$
\ForAll{$\gamma\in\mathcal{U}$}
\If{$\gamma\neq \beta$}
\State $k_{\beta}^{\gamma,t}\leftarrow f_{KnowledgeDecay}(k_{\beta}^{\gamma,t})$
\EndIf
\EndFor
\EndFor
\EndFor
\end{algorithmic}
\caption{Passing of Time Function}\label{alg:passing}
\end{small}
\end{algorithm}

Note that the passing of time function can be executed on the background by an IAA at regular time intervals without delaying its user's actions. 

\subsection{Message Reception} The pseudo code corresponding to this function is depicted in Algorithm \ref{alg:reception}. At this step, an IAA processes the messages received by its user. Messages are received in the input mailbox ($InBox$). The IAA treats the messages received in the input mailbox in FIFO order for:
\begin{enumerate}
	\item Updating the exchange lists. The IAA computes the appropriateness of all messages its user receives to update the appropriateness exchange list with the sender user of each message (line 3). Similarly, the knowledge of messages is used to update the knowledge exchange list (line 4).
	\item Updating the appropriateness likelihoods. All topics mentioned in a message are used to update the likelihood that the sender approves to exchange information about these topics (line 6).
	\item Updating the knowledge likelihoods. All topics exchanged in a message are used to update the likelihood that the association of the tagged users with the topics is known by the sender (line 8). Similarly, 
the likelihood that the association of the tagged users with the topics is known by the rest of message receivers is increased as well (line 10).
\end{enumerate}

\begin{algorithm}
\begin{small}
\begin{algorithmic}[1]
\While {$InBox$ is not empty}
\State Retrieve $\langle \beta,R,T,S\rangle$ from $InBox$
\State $\mathcal{A}_{\beta}\leftarrow\mathcal{A}_{\beta}\cup \{f_{Appropriateness}(\langle \beta,R,T,S\rangle))\}$
\State $\mathcal{K}_{\beta}\leftarrow\mathcal{K}_{\beta}\cup \{f_{Knowledge}(\langle \beta,R,T,S\rangle)\}$
\ForAll{$t\in T$}
\State $a_{\beta}^{t}\leftarrow f_{AppropiatenessIncrease}(a_{\beta}^{t})$
\ForAll{$s\in S$}
\State $k_{\beta}^{s,t}\leftarrow f_{KnowedgeIncrease}(k_{\beta}^{s,t})$
\ForAll{$r\in R$}
\State $k_{r}^{s,t}\leftarrow f_{KnowledgeIncrease}(k_{r}^{s,t})$
\EndFor
\EndFor
\EndFor
\EndWhile
\end{algorithmic}
\caption{Message Reception Function}\label{alg:reception}
\end{small}
\end{algorithm}

The message reception function is executed every time a new message is received. In the worst case, the temporal cost of this function is $O(T  F^2)$, where $T$ is the number of topics and $F$ is the number of friends in the social network. Among adult Facebook users the median number of friends\footnote{Source: \url{http://www.pewresearch.org/fact-tank/2014/02/03/6-new-facts-about-facebook/}} is 200. For example, works on NLP state that there are 36 key concepts on written English {Rayson}. According to recent statistics, the average social network user receives 285 pieces of content daily\footnote{Source: \url{http://www.iacpsocialmedia.org/Resources/FunFacts.aspx}}. Thus, on average the IAA executes the message reception function, which has a polynomial cost, less than $300$ times a day. Besides that, that this function does not require the intervention of users. That is, the messages received by an IAA can be straightforwardly delivered to its user while the IAA executes the message reception function  on the background. 

\subsection{Message Sending} The pseudo code corresponding to this function is depicted in Algorithm \ref{alg:sending}. At this step, the IAA processes the messages that its user wants to send before they are actually sent. The user leaves the messages she wants to send in the output mailbox ($OutBox$). Before these messages are sent, the IAA analyses the messages and tries to prevent their users from sending messages in which inappropriate information would be exchanged and/or unknown information could be disseminated. In particular, the IAAs analyses the messages to be sent to:

\begin{enumerate}
	\item Check the appropriateness of messages, which is computed for each message (line 4). The value of the message appropriateness models the suitability of a message with respect to the information sharing norms. Besides, and as aforementioned, people usually reveal inappropriate information (e.g., embarrassing information) to create and/or to reciprocate close relationships. Thus, the appropriateness of the messages that have been exchanged with each receiver is computed as well (line 4). This value models the level of closeness of the relationship between the sender and the receiver. To compute the appropriateness of the messages that have been previously exchanged with each receiver, the IAA uses $f_{Aggregate}:[0,1]^{n}\rightarrow[0,1]$, which is a function that maps a set of real numbers within the $[0,1]$ interval to a real number within the $[0,1]$ interval. This function aggregates all the appropriateness likelihoods (vs. knowledge likelihoods) stored in the exchange list with a receiver user $\mathcal{A}_{r}$ into a single value.
	After this, the IAA checks the appropriateness of the message being analysed with respect to the appropriateness of the messages that have been previously exchanged with the receiver. If the appropriateness of a message is less than the appropriateness of the messages that have been exchanged with the receiver, then the IAA alerts its user about the risks of such action and asks its user to confirm the sending operation\footnote{For the sake of clarity and simplicity, we only show in this paper where the alerts would be raised, but please note that the alerts could be much more informative than what is shown in Algorithm \ref{alg:sending}. For instance, an IAA could use all the information it has about the receivers, information appropriateness, and previous messages exchanged to craft a more informative alert to be conveyed to its user.}. 
	\item Avoid undesired disseminations. An IAA also tries to avoid that its user sends messages that may lead to the dissemination of sensitive and unknown information. Specifically, the knowledge of messages is also computed for each message (line 11). The value of the message knowledge models the sensitivity of a message. The lower the knowledge of a message, the higher the probability it contains sensitive information whose dissemination is undesirable. People usually share unknown information to create or maintain trust relationships  (e.g., I would reveal my husband confidential information about a common friend). The knowledge of the messages that have been exchanged with each receiver is also computed (line 11). When the knowledge of a message is less than the knowledge of the messages that have been exchanged with the receivers, then the IAA alerts its user about the risks of such action and asks its user to confirm the sending operation. 
	\item Update the exchange lists. If the message is finally sent, then the IAA computes the appropriateness of the messages sent by the user to update the appropriateness exchange list with the receiver users (line 18). Similarly, the knowledge of messages is used to update the knowledge exchange list (line 19).
	\item Update the knowledge likelihoods. If the message is finally sent, then all topics mentioned in a message are used by the IAA to update the likelihoods that the association of the tagged users with topics mentioned in the message is known by the receivers (line 22). 
\end{enumerate}

\begin{algorithm}
\begin{small}
\begin{algorithmic}[1]
\While {$OutBox$ is not empty}
\State Retrieve $\langle \alpha,R,T,S\rangle$ from $OutBox$
\For{$r \in R$}
\If{$f_{Appropiateness}(\langle \alpha,R,T,S\rangle)<f_{Agreggate}(\mathcal{A}_{r})$}
\State $continue\leftarrow receiveDecision$(`This may be inappropiate')
\If{$\lnot continue$}
\Return
\EndIf
\EndIf
\EndFor
\For{$r \in R$}
\If{$f_{Knowledge}(\langle \alpha,R,T,S\rangle)<f_{Agreggate}(\mathcal{K}_{r})$}
\State $continue\leftarrow receiveDecision$(`This may disseminate sensitive information')
\If{$\lnot continue$}
\Return
\EndIf
\EndIf
\EndFor
\For{$r \in R$}
\State $\mathcal{A}_{r}\leftarrow\mathcal{A}_{r}\cup \{f_{Appropiateness}(\langle \alpha,R,T,S\rangle)\}$
\State $\mathcal{K}_{r}\leftarrow\mathcal{K}_{r}\cup \{f_{Knowledge}(\langle \alpha,R,T,S\rangle)\}$
\ForAll{$t\in T$}
\ForAll{$s\in S$}
\State $k_{r}^{s,t}\leftarrow f_{KnowledgeIncrease}(k_{r}^{s,t})$
\EndFor
\EndFor
\EndFor
\State Send $\langle \alpha,R,T,S\rangle$
\EndWhile
\end{algorithmic}
\caption{Message Sending Function}\label{alg:sending}
\end{small}
\end{algorithm}

The message sending function is executed before a new message is actually sent. This is the only function that may require the intervention of users by asking them to confirm the sending operation. That is, if the IAA raises an alert, its user needs to confirm whether the message should finally be sent or not. Otherwise, the message will be sent straightforwardly. In the worst case, the temporal cost of this function is $O(TF^2)$, where $T$ is the number of topics and $F$ is the number of friends in the social network. Recent studies on sharing behaviour on social network, showed that users users send messages less frequently than they receive messages\footnote{Source: \url{http://www.pewinternet.org/files/2015/01/PI\_SocialMediaUpdate20144.pdf}}. Thus, the message sending function will be executed significantly less times than the message reception function.

As aforementioned, there is a single IAA situated between the user and the ONS. However, it may be possible to distribute the computation performed by an IAA into a set of specialized IAAs. For example, the messages sent and received can be distributed among specialized IAAs according to the context. Thus, each IAA is responsible for controlling the interactions of a user within a given context. Finally, it is also possible to dynamically adapt the number of IAAs by performing cloning and self-deletion operations. However, the definition of more elaborated procedures for adapting dynamically to changing environments \cite{nakano2005self} is a complex issue that is out the scope of this paper.

\section{Illustrative Examples}\label{sec:examples}
This section illustrates the performance of the IAA in a few representative examples.  In particular, these examples  are aimed at showing the capabilities of IAAs to deal with: (i) evolving relationships between users; (ii) dynamic contexts; and (iii) distinguishing unusual information from inappropriate information.  

\subsection{Evolving Relationships}
To illustrate the performance of IAAs when dealing with evolving relationships between users, let us assume the existence of a user $\alpha$ who has just started working at a new company. $\alpha$ uses the IAAs to interact with his workmates, represented as the context $ C_{work}=\{\beta,\gamma,\delta\}$, through an OSN. Initially, none message has been exchanged between them. As a result, all the individual appropriateness values are set to 0.5 (i.e., $\forall c\in C_{work},t\in \mathcal{T}: a_c^t=0.5$). Similarly, the appropriateness exchange lists are empty (i.e., $\forall c\in C_{work}: \mathcal{A}_c=\emptyset$). From that moment on, several messages containing the $work$ topic (i.e., messages about tasks, meetings, etc.) are exchanged between them. Thus, the $work$ topic is seen by all users as appropriate (i.e., $\forall c\in C: a_c^{work}\approx 1$), whereas the rest of topics are seen as inappropriate (i.e., $\forall c\in C_{work}, t\in \mathcal{T}: t\neq work \wedge a_c^{t}\approx 0$). Similarly, the appropriateness exchange lists contain a list of increasing values (i.e., $\forall c\in C_{work}:\mathcal{A}_c\approx\{0.5,0.7,\ldots,1,1,1\}$), which represents the fact that the work topic has become more appropriate to the group as they have advanced in their professional relationship.

At some point, $\alpha$ and $\beta$ realised they both support Manchester United and they begin to exchange some football videos between them. Obviously this causes that the IAA infers that $\beta$ considers as appropriate to exchange information about the $sport$ topic (i.e., $a_\beta ^{sport}\approx 1$). However, the IAA infers that among the work friends the $sport$ topic is mostly inappropriate (i.e., $f_{ContextAppropiateness}(C_{work},\{sport\})\ll 1$), since the rest of friends never mentioned it. This causes that the IAA infers that $\beta $ has become a close colleague and the relationship between them has evolved (i.e., $\mathcal{A}_\beta=\{\ldots,1,\ldots,0.3,0.3\}$), since there are some common interests that lead to messages including inappropriate topics. Because of that, when $\alpha $ sends a message to $\beta$ with the $sport$ topic no alert is raised (note that the IAAs knows that the $sport$ topic is inappropriate for the work context, but $\alpha$ and $\beta $ usually exchange information that is not appropriate for this context). 

Sometime afterwards, $\beta $ becomes $\alpha $ boss and he thinks that he needs to have a more professional relationship with his workmates (the relationship evolves again). As a result, $\beta$ stops sending messages about the $sport$ topic, and only exchanges work-related information with $\alpha$. As a consequence, the appropriateness exchange list reflects that lately $\beta$ has only being exchanging appropriate information (i.e., $\mathcal{A}_\beta=\{\ldots,1,1,1\}$). Thus, the next time $\alpha$ tries to send a message to $\beta$ containing football information, the IAAs will raise an alert indicating that this may been seen as inappropriate. Then, $\alpha$ must decide between sharing the information anyway (trying to rebuild the relationship) or not sharing, thus abiding by the new roles.

\subsection{Evolving Contexts}
\begin{figure}
	\centering
        \begin{subfigure}[b]{0.32\textwidth}
                \includegraphics[trim = 0mm 140mm 110mm 0mm, clip, width=0.8\textwidth,page=1]{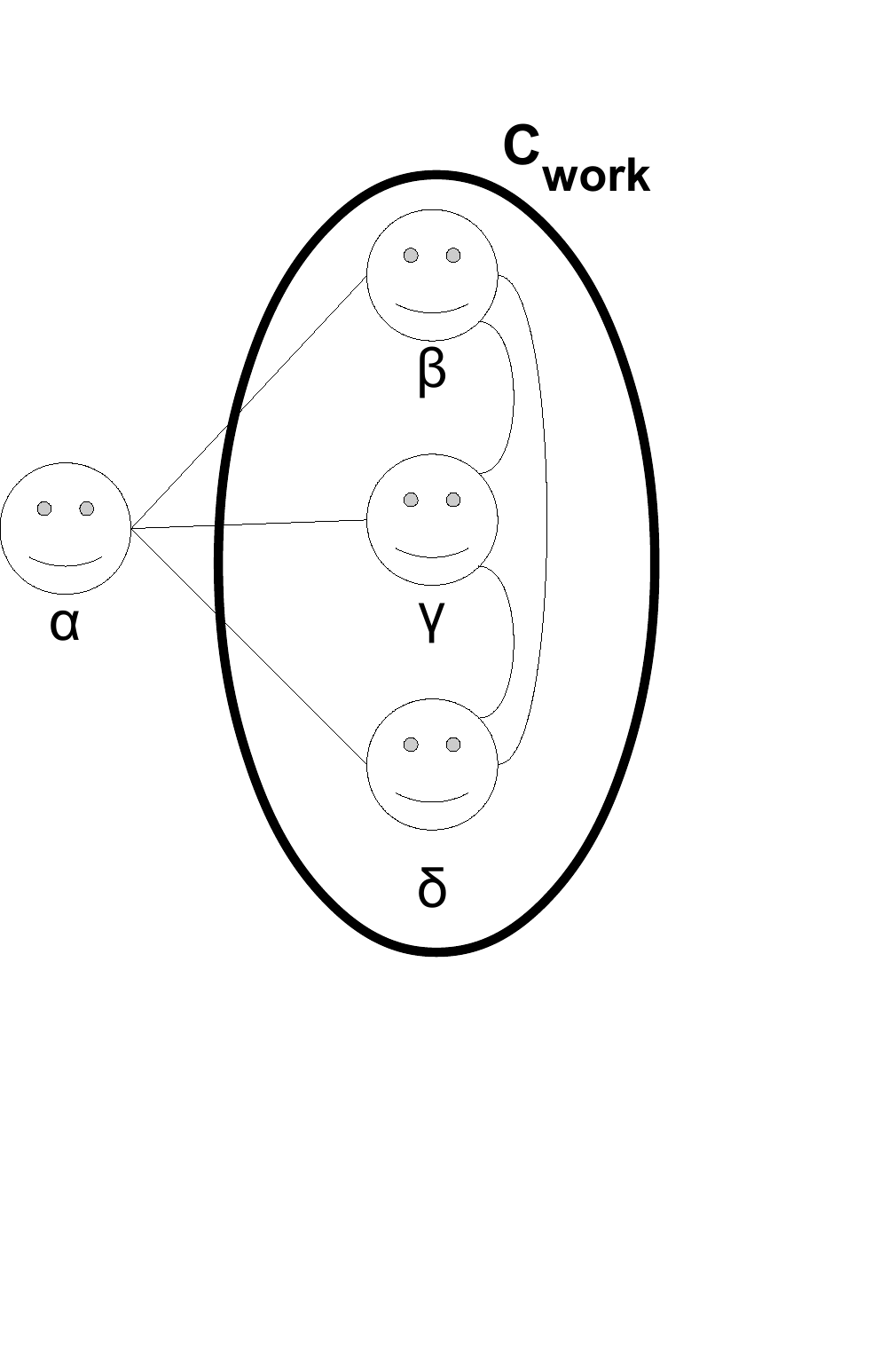}
\vspace{-10pt}
                \caption{Initial Situation}
                \label{fig:initial}
        \end{subfigure}
        \begin{subfigure}[b]{0.32\textwidth}
                \includegraphics[trim = 0mm 140mm 110mm 0mm, clip, width=0.8\textwidth,page=2]{Example.pdf}
\vspace{-10pt}
                \caption{First Photography Class}
                \label{fig:first}
        \end{subfigure}
        \begin{subfigure}[b]{0.32\textwidth}
                \includegraphics[trim = 0mm 90mm 110mm 0mm, clip, width=0.8\textwidth,page=3]{Example.pdf}
                \caption{Second Photography Class}
                \label{fig:second}
        \end{subfigure}
        
\vspace{-5pt}
	\caption{Friend Network in the Evolving Context Scenario. Users are represented by smiley faces, friend relationships between users are represented by lines, and contexts are represented by bold ellipses. }
	\label{fig:ExampleScenario}
\end{figure}

In this example, we will focus on a situation in which contexts evolve. Let us assume the same user $\alpha$ with the same three workmates $\beta,\gamma$ and $\delta$. $\alpha$ has not defined any group on the OSN and the IAA uses the community finding algorithm to infer that $\beta,\gamma$ and $\delta$ belong to the same context (i.e., $C_{work}=\{\beta,\gamma,\delta\}$). The social network corresponding to this situation is depicted in Figure \ref{fig:initial}. 

At some point, $\alpha$ and $\delta$ start a weekly photography course. In the first class, they meet $\epsilon$ and become his friends on the social network. This situation is represented in Figure \ref{fig:first}. As a result of this change in $\alpha$'s friend network, the IAA executes the community finding algorithm. In the absence of more information about user-generated groups, the algorithm assigns $\epsilon$ to the work context (i.e., $C_{work}=\{\beta,\gamma,\delta,\epsilon\}$).  

In the second week, $\delta$ decides to leave the course whereas $\alpha$ and $\epsilon$ attend the second class. In that class, they meet $\zeta$ and $\eta$ and become their friends on the social network. This situation is represented in Figure \ref{fig:second}. As a result of this change in $\alpha$'s friend network, the IAA executes the community finding algorithm. Thanks to the new friend relationships, the algorithm identifies correctly the work context (i.e., $C_{work}=\{\beta,\gamma,\delta\}$) and the photography course context (i.e., $C_{course}=\{\epsilon,\zeta,\eta\}$).  

This example illustrates how IAAs are able to deal  with the dynamic creation of unforeseen contexts. 

\subsection{Distinguishing Inappropriate Information from Unusual Information}
As previously mentioned, the likelihood values of the proposed information model are estimated as frequentist probabilities. This entails that topics that are frequently discussed by users in a context are seen as appropriate. Topics that are never mentioned may be inappropriate or unusual. This example illustrates how IAAs distinguish between inappropriate and unusual information. 

Let us assume the same users $\alpha,\beta,\gamma$ and $\delta$. As previously mentioned, these users use the OSN to exchange work-related information. Thus, the topic $work$ is considered as appropriate by the IAAs. Any time $\alpha$ decides to send a message discussing work-related issues, the IAA processes the message without raising any alert. At some point, $\alpha$ sends a message containing an invitation to his wedding. The IAA raises an alert informing $\alpha$ about the fact that the message may be seen as inappropriate (i.e., $f_{ContextAppropiateness}(C_{work},\{marriage\})=0$ since $\forall c\in C_{work}: a_c^{marriage}=0$). $\alpha$ confirms the sending operation. As a result, $\beta,\gamma$ and $\delta$ express their joy (e.g., they like the original post) and reply to the original post (e.g., they make comments on the post indicating whether or not they will attend the wedding). All these messages entail that the IAA increases the appropriateness of the $marriage$ topic (i.e., $\forall c\in C_{work}: a_c^{marriage}\gg 0$). When $\alpha$ decides to send other messages with further details about the wedding, the IAA does not raise any alert.

Let us assume a different situation in which $\alpha$ decides to send a message containing an obscene joke. Given that the $obscene$ topic has never been mentioned before, the IAA raises an alert (i.e., $f_{ContextAppropiateness}(C_{work},\{obscene\})=0$ since $\forall c\in C_{work}: a_c^{obscene}=0$). However, $\alpha$ decides to send the message anyway. However, $\beta,\gamma$ and $\delta$ think that the joke is in bad taste and they ignore it (e.g., they do not comment it , like it or retweet it). As a result, the appropriateness of the $obscene$ topic does not increase (i.e., $\forall c\in C_{work}: a_c^{obscene}=0$). When $\alpha$ decides to send another message containing an obscene joke the IAA raises an alert. At this point, $\alpha$ might realise that their friends do not approve this kind of information and decide to stop sending these messages.

This example shows how IAAs can distinguish between inappropriate and unusual information. In both cases the IAA raises an alert when the first message is sent. However, the IAA only continues to raise alerts when the information is inappropriate (e.g., obscene jokes), not when it is unusual (e.g., wedding  details). In particular, the fact that users of OSNs usually show their support for specific comments, pictures, posts, etc. can be exploited to distinguish unusual form inappropriate information. Indeed, recent statistics show that this form of communication (i.e., likes and comments) is prevalent in OSNs\footnote{According to Instagram statistics, on average 2.5 billion likes are sent diary, whereas only 70 million photos are uploaded per day. Source: \url{https://instagram.com/press/}}. 

\section{Evaluation}\label{sec:evaluation}
To analyse how the IAAs works and to validate our hypothesis about their behaviour, this section analyses the properties of IAAs regarding: contextual integrity maintenance, information sharing norm learning, alert notifications to their users, effect of environmental parameters, and robustness to malicious users. To this aim, we conducted a series of experiments illustrating the behaviour exhibited by IAAs considering a wide range of different situations and conditions, modelling scenarios with real OSN characteristics (e.g., scale-free networks, few strong ties, etc.) and heterogeneous users with different attitudes and preferences.


\subsection{Setting Description}   

We considered users who are members of an OSN and who initially have some pieces of information that can be exchanged. As they receive information from others, they acquire new pieces of information that can also be exchanged. Parameters and their default values (which change depending on the experiment) are described in Table \ref{tab:experiments}. 
Each user on the OSN has been implemented as a \textit{user} agent. There is a set $\mathcal{U}$ of $100$ user agents among which friendship relationships are randomly created following a scale-free pattern, according to empirical studies that prove OSNs are scale-free \cite{mislove2007measurement}. In our scenario, we assume that users can only exchange information with their friends (as occurs in Facebook and other mainstream OSN). A subset of these friendship relationships is randomly selected to be close and trusted friend relationships\footnote{For simplicity, in our simulations we have assumed that all close relationships are also trust relationships. However, our model is able to infer these two types of relationships independently.}. Users are grouped forming contexts or communities. The set of contexts ($\mathcal{C}$) in each simulation (considering the \emph{complete} view of the social network) is elicited using the Infomap algorithm \cite{rosvall2007information}, which is a clustering algorithm that has demonstrated high performance when identifying Facebook contexts \cite{fogues2014bff}. Note that these are unknown to IAAs, which only have an approximation to these contexts in each simulation by running the same algorithm but with a \emph{partial} view of the social network from their users point of view (as stated in Section \ref{sec:agent}). This makes even more difficult for IAAs to learn the information sharing norms, but this models the conditions IAAs will actually face in real OSNs.

\begin{table}
\begin{tabular}{cc}
\hline
Parameter&Default Value\\
\hline
number of user agents &100\\
number of topics &36\\
maximum number of topics per message &36\\
maximum ratio of inappropriate topics in each context &10\%\\
maximum ratio of sensitive associations in each context &1\%\\
number of steps& $2000$\\
number of executions& $100$\\
\hline
\end{tabular}
\caption{Parameters and their default values.}
\label{tab:experiments}
\vspace{-10pt}
\end{table}

Next, we describe how the information sharing norms, users, and IAAs were modelled:

\subsubsection{Information Sharing Norm Modelling}
At the beginning of each simulation, we created a set of information sharing norms, which determined the information that is inappropriate in each context and the information that is sensitive (i.e., mostly unknown) in each context and should not be disseminated. Note that these norms are implicit and a priori unknown by users and agents.

 \begin{definition}
 Formally, the set of information sharing norms is defined as follows:

 \[\mathcal{N}=\mathcal{N}_{Inappropiate}\cup\mathcal{N}_{Sensitive}\]

 where:
 \begin{itemize}
 	\item $\mathcal{N}_{Inappropiate}$ is the set of inappropriateness norms that is formed by $\langle t,c\rangle$ expressions, which represent that talking about topic $t\in\mathcal{T}$ is seen as inappropriate in context $c\in\mathcal{C}$;
 	\item $\mathcal{N}_{Sensitive}$ is the set of sensitiveness norms which is formed by $\langle u,t,c\rangle$ expressions, which represent that the association of user $u\in\mathcal{U}$ with topic $t\in\mathcal{T}$ is sensitive (i.e., mostly unknown) in context $c\in\mathcal{C}$.
 \end{itemize}
 \end{definition}
We considered a set $\mathcal{T}$ of 36 different topics that represent key concepts (e.g., sports, politics, etc.) and that can be recognised in English text \cite{Rayson}. In each context, we define randomly the topics that are inappropriate and associations between users and topics that are sensitive: at maximum 10\% of these topics are inappropriate (we vary this later on), so that the number of inappropriate topics in each context belongs to the interval $[1,4]$; and at maximum 1\% of the possible associations between users and topics are sensitive (we vary this later on), so that the number of sensitive associations in each context belongs to the interval $[1,36]$. 

Social norms, as information sharing norms in OSN, are not created by legal or institutional authorities. They are established in societies by means of a bottom-up process in which they emerge from individual behaviour \cite{elster1989social}. Specifically, social norms emerge when they are followed by a considerable portion of the society. To simulate the existence of the information sharing norms (i.e., the appropriateness and sensitiveness norms IAAs need to learn), we model a subset of the users in the OSN to be \emph{compliant users}, who do not use IAAs but know the norms and follow them; i.e., they simulate people that have the ability to
say always the right thing. Thus, compliant users only send messages that respect the appropriateness and sensitiveness norms.

\subsubsection{User Modelling}
Apart from compliant users, we modelled the behaviour of different types of users according to a classification of different normative attitudes based on Westin's privacy attitudes \cite{kumaraguru05,dma2012}:
\vspace{-3pt}
\begin{itemize}
	\item Random users are norm \textit{unconcerned} --- i.e., users who do not care about the information sharing norms, so they do not use IAAs and randomly compose messages containing information that they know and send them to a random subset of their friends. These messages can cover any set of topics.

	Note that ours is the first model of implicit contextual integrity, thus we cannot compare the performance of IAAs against previous proposals. As a consequence, we will use random users (which model users who do not use our methods to maintain contextual integrity) as a baseline method for comparison, which simulates the current situation in all OSN infrastructures (like Facebook, Google+, etc.).
	\item Obedient users are norm \textit{fundamentalists} --- i.e., users who adhere to the information sharing norms\footnote{Note that compliant users can also be classified as norm fundamentalists.}. They utilise IAAs and are called obedient since they always follow the recommendations made by their IAAs. Similarly to compliant users, they also try to comply with the information sharing norms, but they do not know the norms and rely on IAAs to infer and abide by the norms. 
	\item Relationship-based users are norm \textit{pragmatists} --- i.e., users who make use of casual forms of information sharing that are not always normative, e.g., when they interact with a close or trusted friend. They utilise IAAs but they decide not to follow the recommendations made by IAAs when they send messages to close and trusted friends.
	
\end{itemize}

\subsubsection{IAAs Implementation}
As aforementioned, the information model used by IAAs makes use of several functions (i.e., update functions, combination function and aggregation function). The propose of this paper is not to provide or compare different definitions to these functions, but to make use of these functions in evaluating anew computational model of contextual integrity. The specific definitions may depend not only on the application domain but also on the personality traits of users\footnote{A user study to generate personalised IAAs is beyond the scope of this paper.}. As a result, in our experiments IAAs’ functions are given well-known and reasonable definitions.

In particular, we define $\bigtriangleup$ as a constant update rule; i.e., $\forall \delta\in[0,1]: \bigtriangleup(\delta)=x$ where $x=0.1$. Similarly, we define $\bigtriangledown$ as a constant update rule; i.e., $\forall \delta\in[0,1]: \bigtriangledown(\delta)=y$ where $y=0.01$. Constant update rules, which have been previously used in the literature on norm learning and emergence \cite{brooks2011modeling}, assume that each IAA will update its likelihood values in constant increments or decrements.

$f_{Combine}$ is implemented as a harmonic mean, since it mitigates the influence of large outliers (e.g., close workmates to which inappropriate information is frequently sent). 

$f_{Aggregate}$ is implemented as a weighted mean where each element is weighted by its index; so that the most recent elements contribute more to the final value. Formally:
\vspace{-9pt}

\[f_{Aggregate}(\mathcal{S})=\frac{\sum_{v_{i}\in\mathcal{S}}i\cdot v_{i}}{\sum_{v_{i}\in\mathcal{S}}i}\]


Finally, IAAs initialize randomly the likelihood that other users approve to exchange topics or know the associations of users with topics. This models a situation in which IAAs have varied and imperfect prior knowledge available\footnote{Note that if we model situations in which all users are new to the OSN and IAAs have no prior knowledge available, which is highly unlikely in practice; this may lead to all IAAs having the same initial values and to the emergence of fake norms corresponding to this shared initialization.}. More formally, for all user $\beta \in \mathcal{U}$ and topic $t \in \mathcal{T}$ the value $a_{\beta}^{t}$ is set to a random real value within the $[0,1]$ interval. Similarly, for all pairs of users $\beta,\gamma \in \mathcal{U}$, where $\gamma<>\beta$ and topic $t \in \mathcal{T}$ the value $k_{\beta}^{\gamma,t}$ is set to a random real value within the $[0,1]$ interval.



\subsection{Experiments}
\subsubsection{Contextual Integrity Maintenance}
In our first experiment we sought to determine if IAAs contribute to maintain contextual integrity, i.e., if IAAs help to reduce the exchange of inappropriate information and the dissemination of sensitive information. To this aim, we simulated societies populated with compliant users (who know and follow the norms) and obedient users (who use IAAs to infer and follow the norms). Each simulation was executed during $2000$ steps and was repeated $100$ times to support the findings. For each execution, we calculated the messages that exchange inappropriate information and the messages that disseminate sensitive information. We also repeated this experiment increasing the percentage of compliant users (who know and follow the norms) in each context from 0\% to 90\% to test IAAs in a wide range of situations. The rest of users in each context are obedient users. As a baseline simulation, we repeated this experiment using compliant users increasing from 0\% to 90\% and the rest being random users (who do not make use of IAAs).

\begin{figure}[h!]
\begin{center}
\includegraphics[trim = 15mm 20mm 5mm 20mm, clip, width=0.75\textwidth]{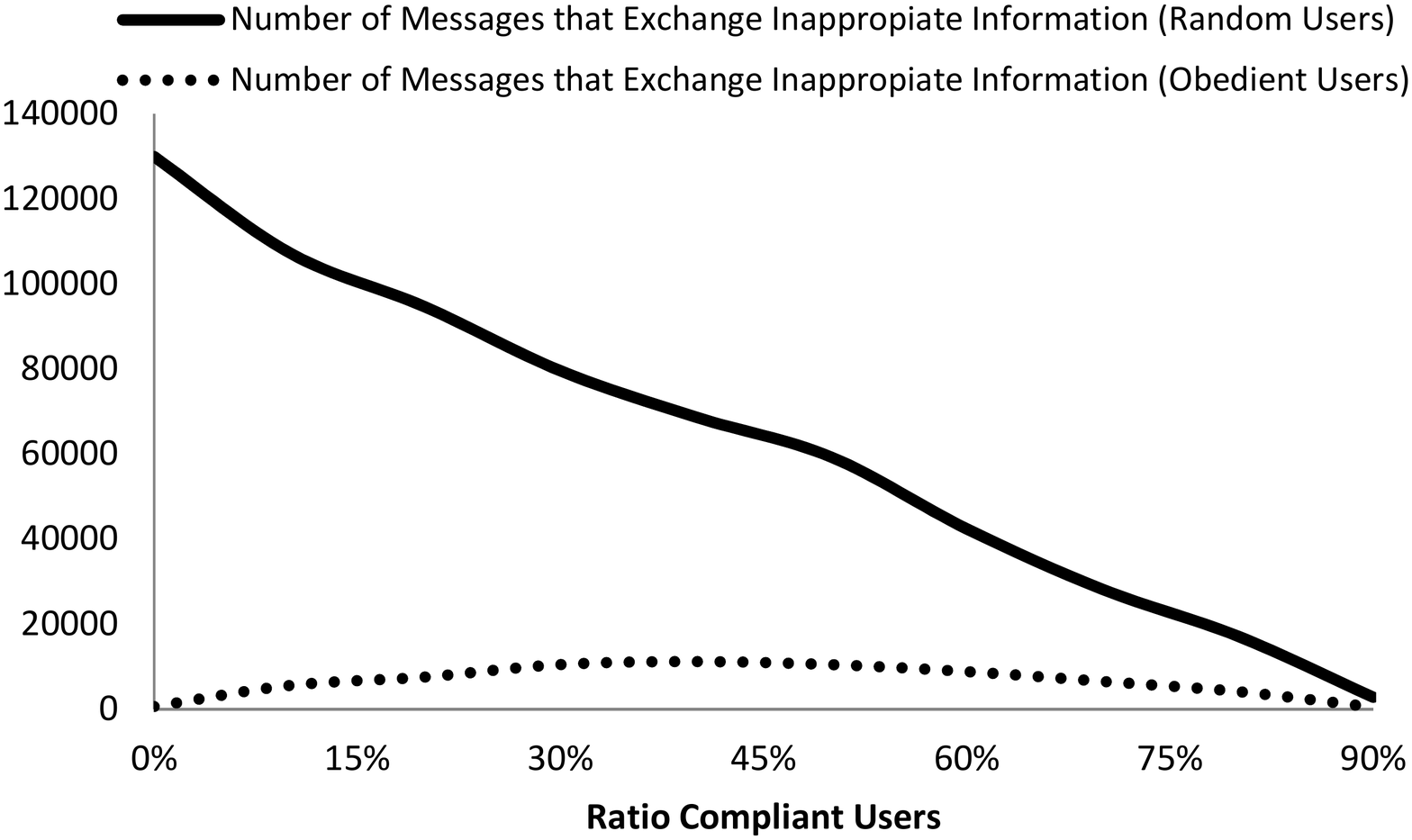} 
\end{center}
\vspace{-10pt}
\caption{Average number of messages that exchange inappropriate information sent by user type (Y-axis) per ratio of compliant users (X-axis)}
\label{fig:explearn1}
\end{figure}

Figure \ref{fig:explearn1} shows the behaviour exhibited by random and obedient users in terms of the exchange of inappropriate information. This figure depicts the average number of messages that exchange inappropriate information sent by all obedient users and by all random users per ratio of compliant agents. This figure shows that obedient users (who make use of IAAs) exchange less inappropriate information than random users (who do not use IAAs) regardless of the ratio of compliant users. On average, the exchange of inappropriate information in societies populated by compliant and obedient users is reduced by $90\%$ compared to societies populated by compliant and random users, which clearly shows that IAAs contribute to maintain contextual integrity. One can also observe that the exchange of inappropriate information in societies populated by random users decreases as the ratio of compliant users increases. Obviously, the less random users, the more compliant users and the less exchanges of inappropriate information occur. This very same effect can be observed in obedient users when the ratio of compliant users belongs to the [$40\%$,$90\%$] interval. As the ratio of compliant users increases more appropriate information is exchanged in the network and obedient users can infer which topics are appropriate and the exchange of inappropriate information decreases more easily. In contrast, when there are few compliant users (i.e., when the ratio of compliant users belongs to the [$0\%$,$40\%$] interval) that exchange appropriate information, then obedient users infer that almost all topics are inappropriate and they end up exchanging less messages\footnote{Note that IAAs are not able to distinguish compliant users and they infer the norms from the information that is exchanged in the social network.}.  

\begin{figure}[h!]
\begin{center}
\includegraphics[trim = 15mm 20mm 5mm 20mm,  clip, width=0.75\textwidth]{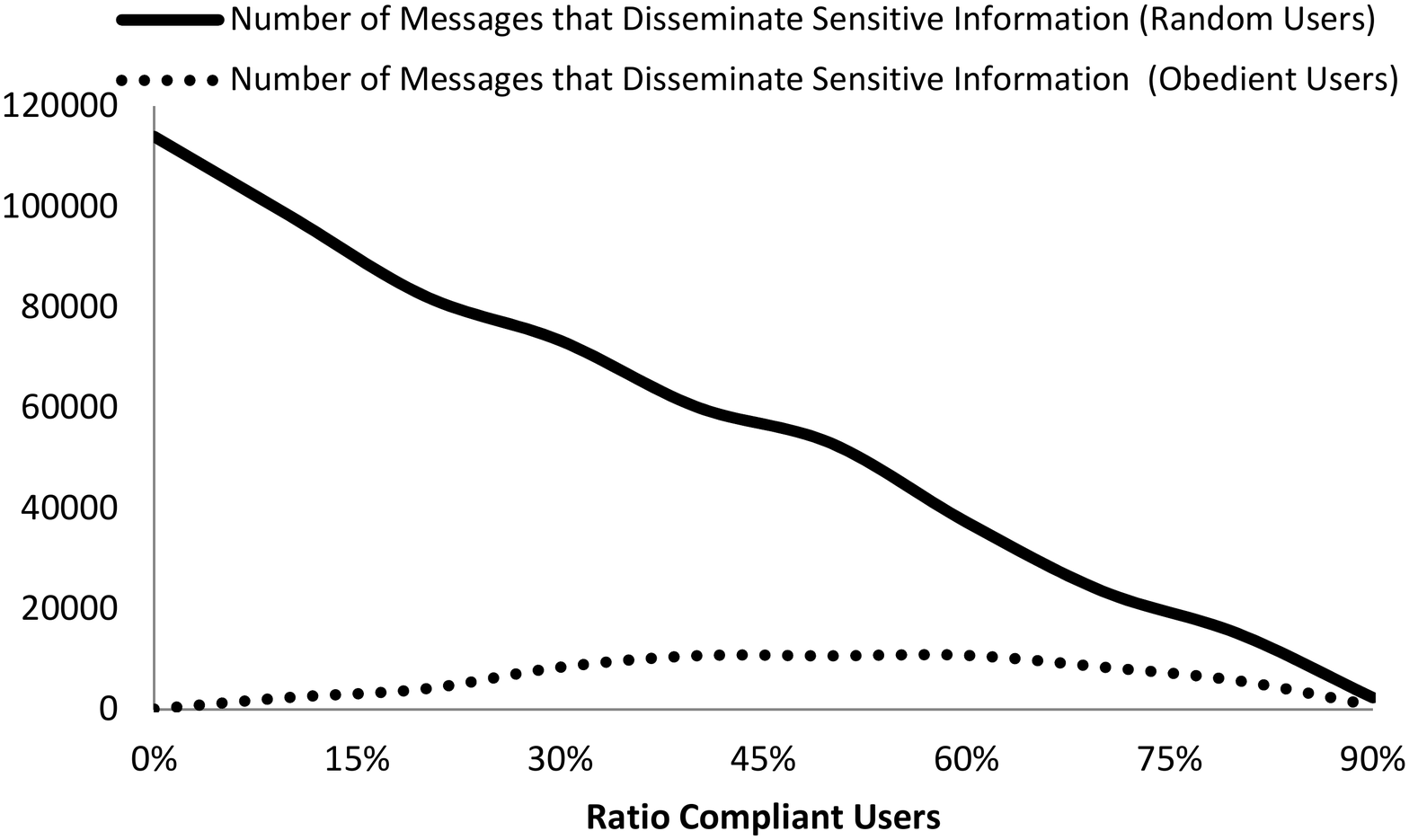} 
\end{center}
\vspace{-10pt}
\caption{Average percentage of messages that disseminate sensitive information sent by user type (Y-axis) per ratio of compliant users (X-axis)}
\label{fig:explearn2}
\end{figure}

Figure \ref{fig:explearn2} compares the behaviour exhibited by random users against obedient users in terms of the dissemination of sensitive information. This figure depicts the average percentage of messages that disseminate inappropriate information sent by all obedient users and by all random users per ratio of compliant agents. As we can observe, the results are very similar to the ones above, i.e., obedient users disseminate less sensitive information regardless of the ratio of compliant users. On average, the dissemination of sensitive information in societies populated by compliant and obedient users is reduced by $89\%$. 

\subsubsection{Information Sharing Norm Learning}
In the previous experiment, we demonstrated that IAAs maintain contextual integrity when the network is formed by compliant users (who know and follow the norms) and obedient users (who try to learn and follow the norms). A more complex scenario arises when IAAs deal with societies in which there are users that do not care about the information sharing norms (i.e., random users). In this situation, it is more difficult to infer the norms on the basis of the information exchanged, since random users exchange appropriate and inappropriate information indistinctly. To assess to what extent IAAs are able to learn the information sharing norms in this kind of situations, we performed an experiment in which we increased the percentage of compliant users in each context from 0\% to 90\%. The rest of users in each context was split in 10\% of obedient users and the rest being random users. Thus, we sought to determine if IAAs are able to infer the norms when they are not predominant in the society. Again, the simulation was executed during $2000$ steps and repeated $100$ times. For each execution, we calculated the messages that exchange inappropriate information and the messages that disseminate sensitive information. 

\begin{figure}[h!]
\begin{center}
\includegraphics[trim = 15mm 20mm 5mm 20mm,  clip, width=0.75\textwidth]{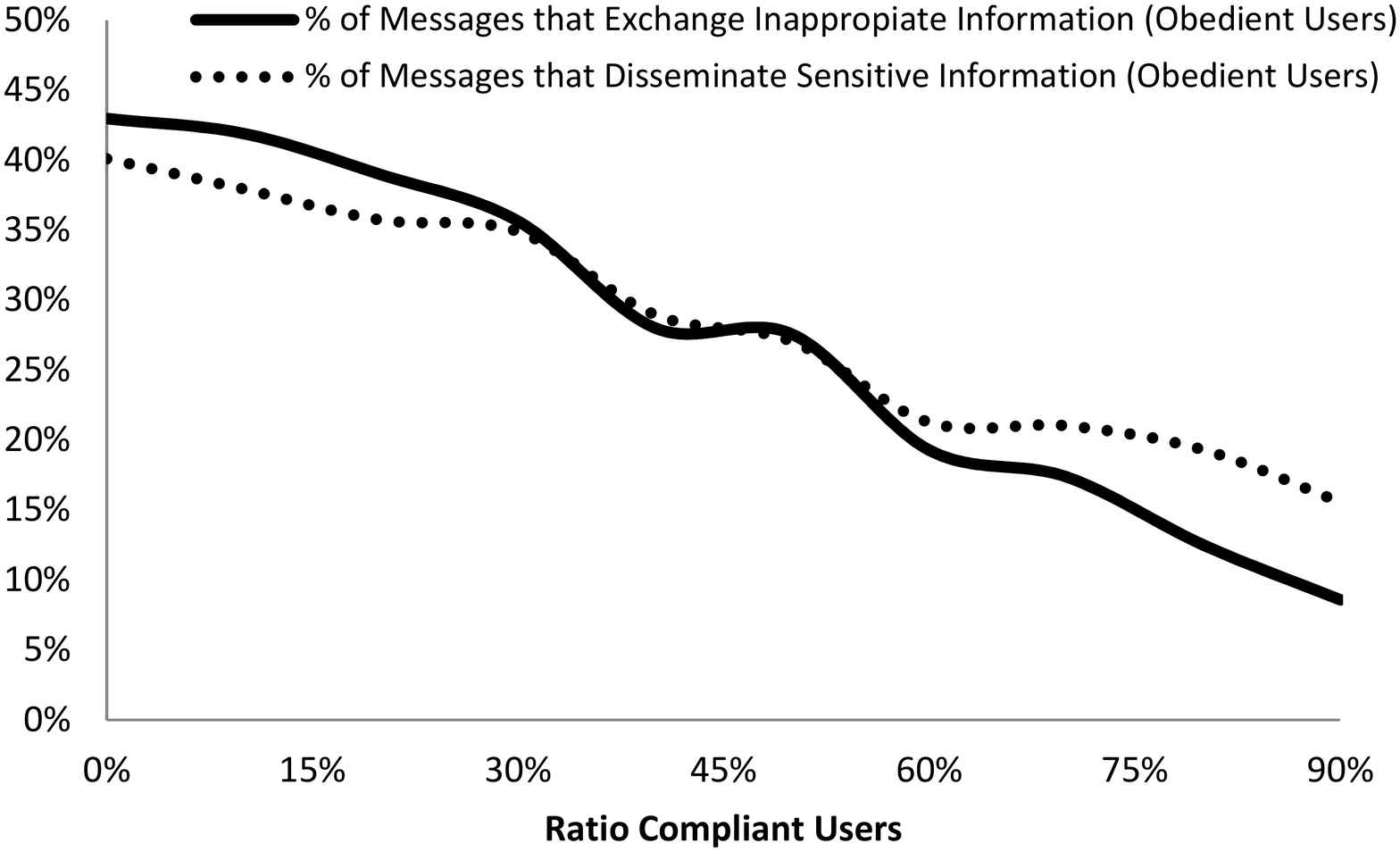} 
\end{center}
\vspace{-10pt}
\caption{Average percentage of messages sent by obedient users that exchange inappropriate information and disseminate sensitive information (Y-axis) per ratio of compliant users (X-axis)}
\label{fig:exp1}
\end{figure}

Figure \ref{fig:exp1} displays the average percentage of messages sent by obedient users (who make use of the IAAs) that exchange inappropriate information and disseminate sensitive information per ratio of compliant agents. This figure shows that the exchange of inappropriate information and the dissemination of sensitive information are reduced noticeable as the ratio of compliant users increases. For example, when there are $40\%$ of compliant users (which implies that there are $10\%$ of obedient and $50\%$ of random users) in each context the exchange of inappropriate information is reduced by $35\%$ and the dissemination of sensitive information is reduced by $27\%$. This demonstrates that IAAs learn the information sharing norms even if compliant users are not predominant in contexts. With regard to disseminations, we can observe that in general less sensitive disseminations occur. This is explained by the fact that sensitivity norms relate a specific user with a specific topic and context, whereas inappropriateness norms only relate a topic with a context. Thus, sensitive information is scarcer than inappropriate information. When the ratio of compliant users is very high, then there are more sensitive disseminations than inappropriate exchanges. This is explained by the fact that sensitiveness norms are more difficult to infer (since they relate a specific user and topic). As a consequence, the reduction on the exchange of inappropriate information as the ratio of compliant users increases is less pronounced.

\subsubsection{Alerts Raised}
This experiment is aimed at assessing the performance of IAAs in terms of the number of alerts raised and how many of these alerts are not followed by users. To this aim, we executed simulations populated with compliant users and relationship-based users (who do not follow the recommendations of IAAs when they exchange information with close and trusted friends). We set the percentage of compliant users in each context to 40\% to model already emerged information sharing norms, and we increased the number of steps that of the simulation to determine the number of steps IAAs require to infer the information sharing norms as well as close and trust relationships. Note that learning close and trust relationships allow IAAs to avoid raising alerts when violations of the information sharing norms are explained because the recipients are close or trusted friends. These alerts may be considered as unnecessary and disturbing since users are unlikely to follow them. 

\begin{figure}[h!]
\begin{center}
\includegraphics[trim = 15mm 20mm 5mm 20mm,  clip,width=0.75\textwidth]{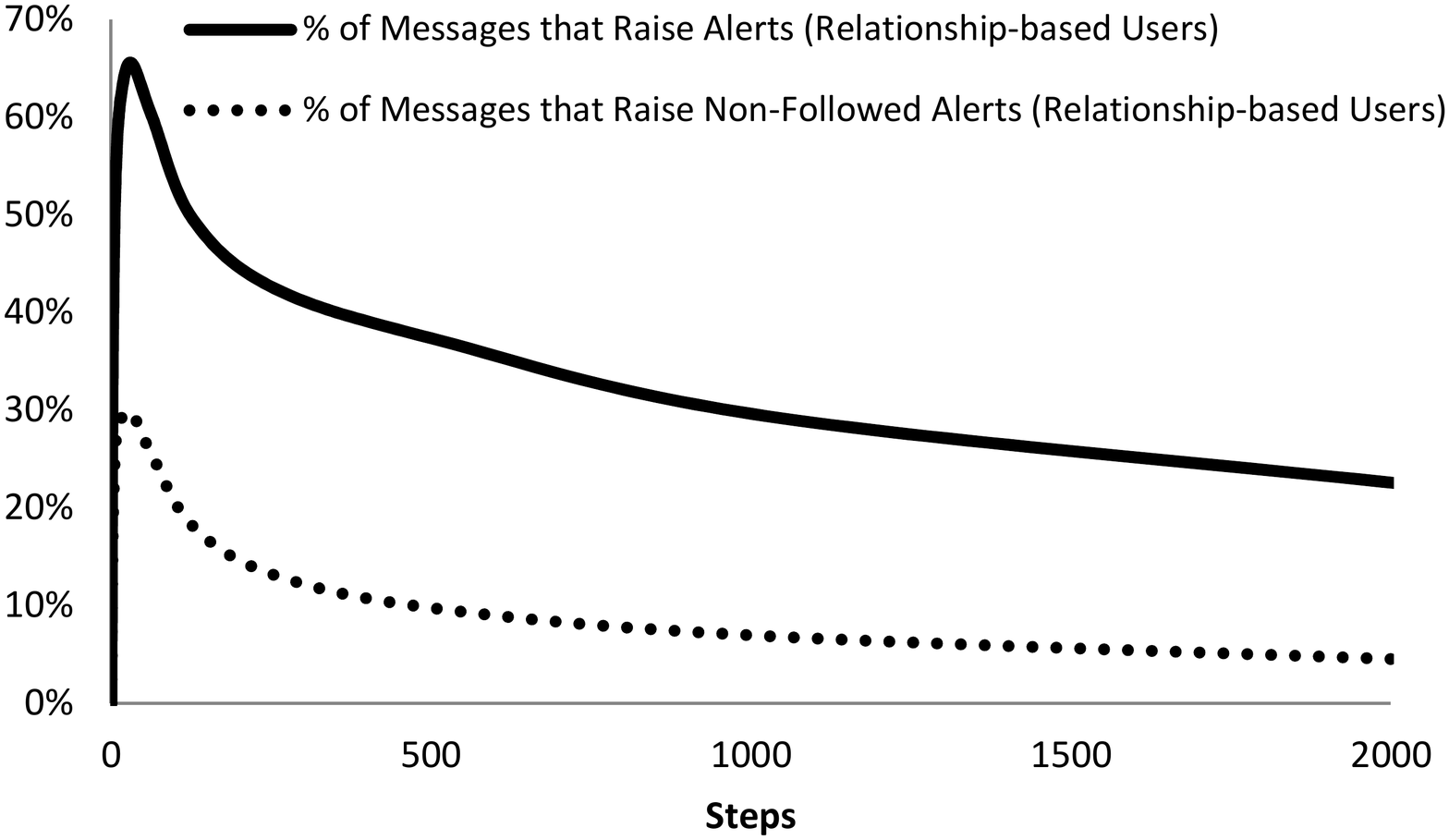} 
\end{center}
\vspace{-10pt}
\caption{Average percentage of raised alerts and average percentage of non-followed alerts (Y-axis) per number of steps (X-axis)}
\label{fig:exp6}
\end{figure}

Figure \ref{fig:exp6} shows that the percentage of messages that raise alerts and the percentage of messages that raise non-followed alerts decrease exponentially to a low value as the number of steps increases, i.e., the percentage of messages that raise alerts converges to a value below $25\%$ and the percentage of messages that raise non-followed alerts converges to a value below $5\%$. This demonstrates that IAAs require a low number of steps to infer the information sharing norms, and trust and close relationships. Also, the low percentage of non-followed alerts ($5\%$) demonstrates that IAAs minimise the burden on the users in terms of raising alerts that would not be followed.

\subsubsection{Simulation Parameters}
In the previous experiments, we fixed the maximum number of topics discussed in a message and the maximum ratios of inappropriate topics and sensitive associations. To analyse the influence of these parameters on the performance of IAAs, we carried out two experiments in which we fixed the percentage of compliant users in each context to 40\% to model already emerged information sharing norms and the rest of users in each context are obedient users. We repeated these two experiments using random users instead of obedient users as a baseline simulation. Each simulation was executed during 2000 steps and repeated 100 times. 

\paragraph{Message Composition}
In our previous experiments, we wanted to assess the performance of IAAs in a general situation; i.e., when users are allowed to select any number of topics to be discussed in a message. Thus, we did not impose any limit on the size of messages. To analyse the impact of the maximum number of topics discussed in the messages (i.e., the message size) on the IAAs performance, we carried out an experiment in which we increased the maximum number of topics that can be discussed in a message from 1 to 35.

\begin{figure}[h!]
\begin{center}
\includegraphics[trim = 15mm 20mm 5mm 20mm,  clip,width=0.75\textwidth]{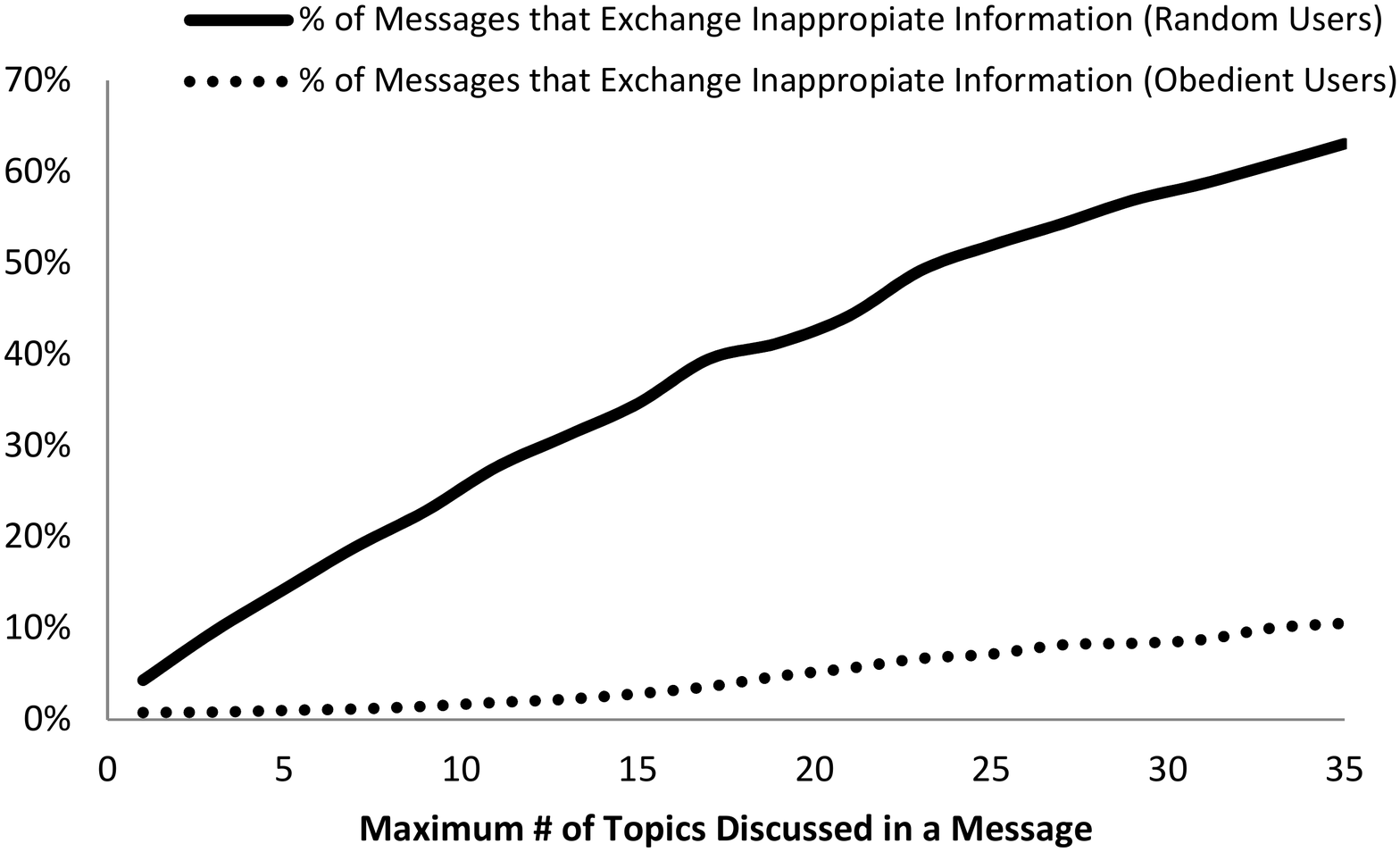} 
\end{center}
\vspace{-10pt}
\caption{Average percentage of messages sent by user type that exchange inappropriate information (Y-axis) per maximum number of topics  (X-axis)}
\label{fig:topicsI}
\end{figure}

\begin{figure}[h!]
\begin{center}
\includegraphics[trim = 15mm 20mm 5mm 20mm,  clip,width=0.75\textwidth]{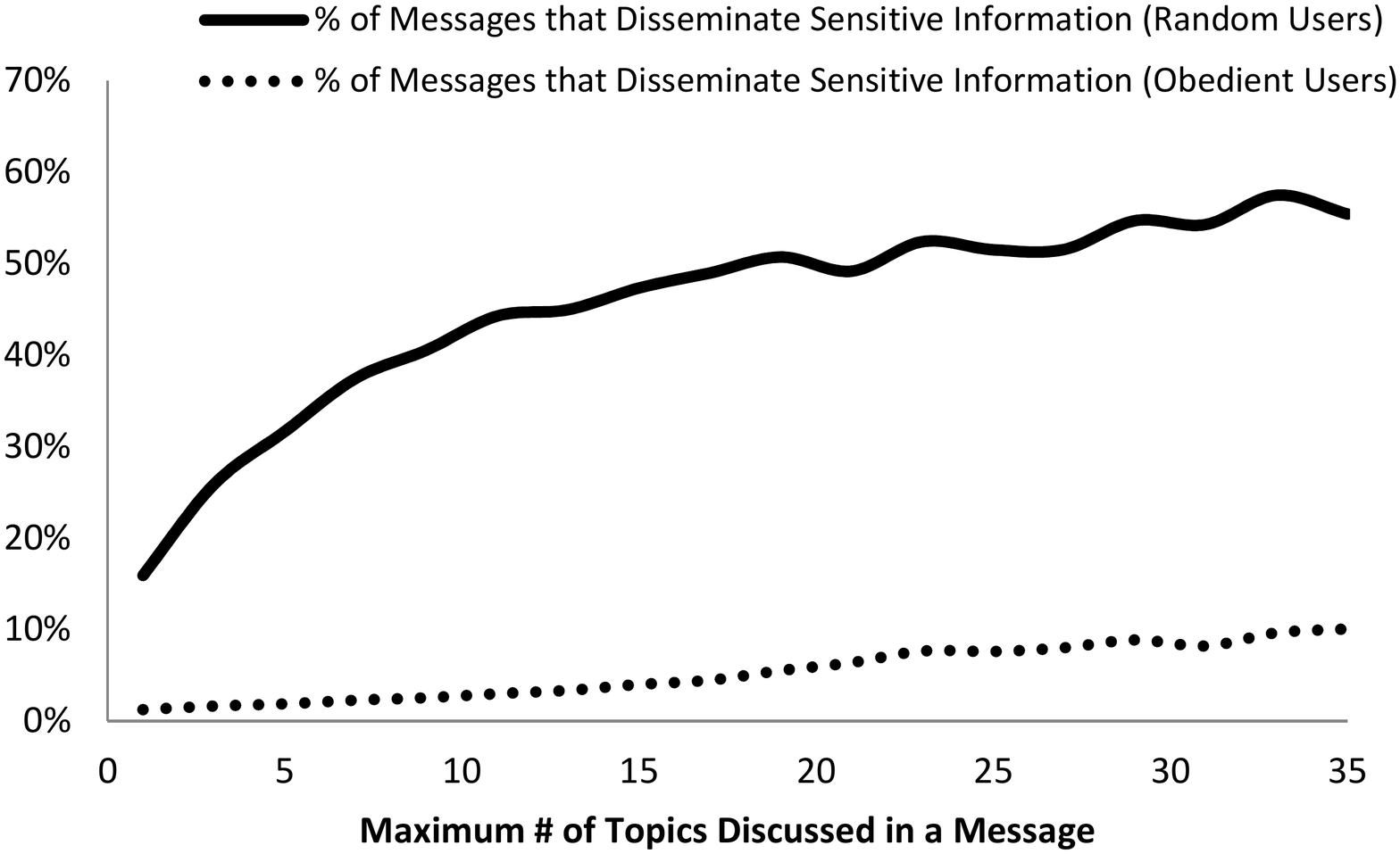} 
\end{center}
\vspace{-10pt}
\caption{Average percentage of messages sent by user type that disseminate sensitive information (Y-axis) per maximum number of topics  (X-axis)}
\label{fig:topicsS}
\end{figure}

Figures \ref{fig:topicsI} and \ref{fig:topicsS} compare the behaviour exhibited by random users against obedient users in terms of the exchange of inappropriate information and the dissemination of sensitive information with respect to the maximum number of topics that can be discussed in a message. As depicted in these two images, the more the topics discussed in messages, the higher the average percentage of messages that violate some information sharing norm. Obviously, the more topics are discussed in a message, the higher the probability of including an inappropriate or sensitive topic. However, the increase on the average percentage of messages that violate some information sharing norm is noticeably higher in random users --- i.e., the lines corresponding to random users have a significantly higher slope than the lines corresponding to obedient users. This demonstrates that even if the users are not able to discern norm-compliant topics to be discussed in messages, IAAs would help them do so.

\paragraph{Norm Definition}
To analyse the impact of the number of norms on IAAs' performance, we carried out an experiment increasing the maximum ratio of inappropriate topics in any context from $5\%$ to $30\%$ and the maximum ratio of sensitive associations between
users and topics from $0.5\%$ to $3\%$.

\begin{figure}[h!]
\begin{center}
\includegraphics[trim = 15mm 20mm 5mm 20mm,  clip,width=0.75\textwidth]{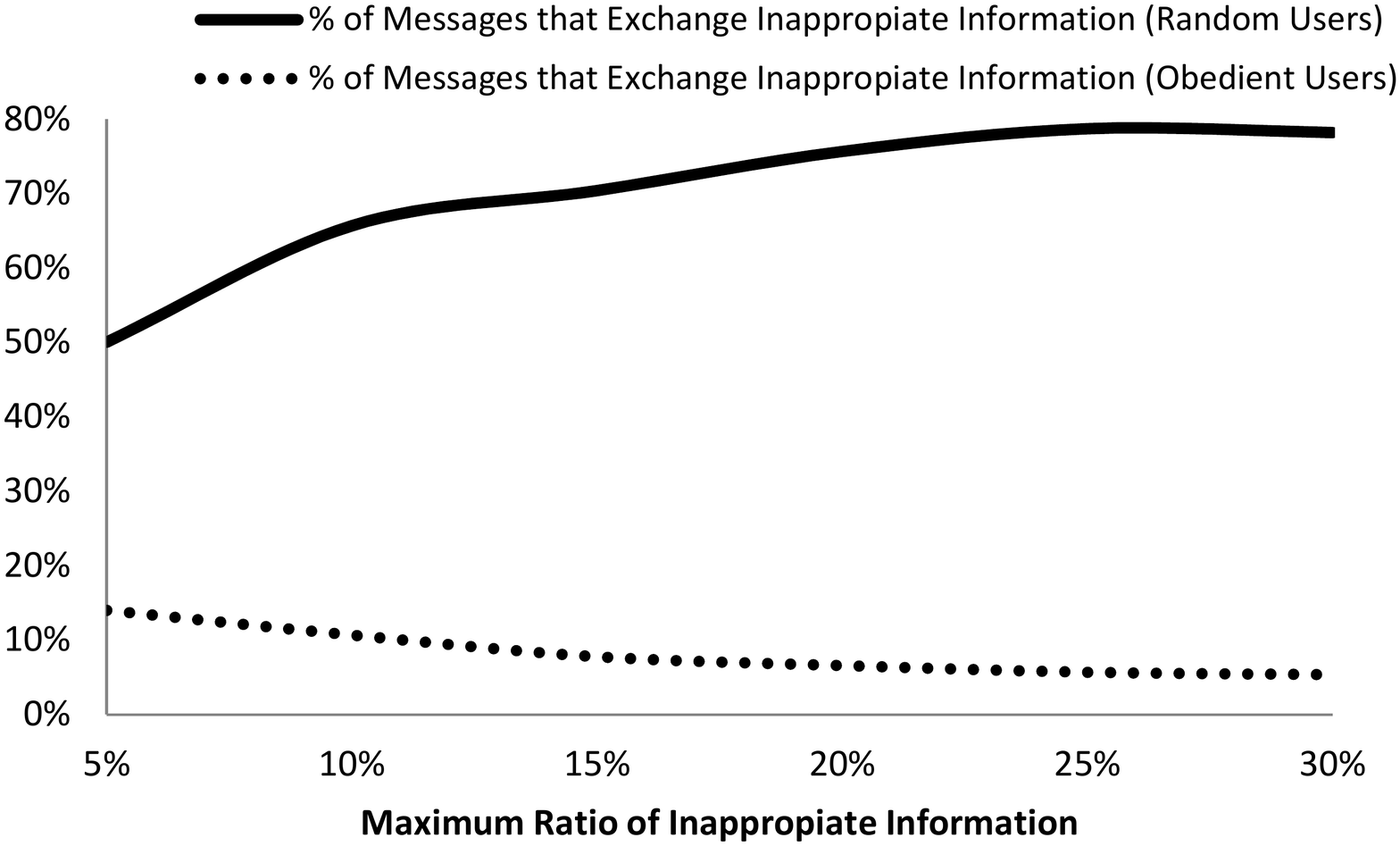} 
\end{center}
\vspace{-10pt}
\caption{Average percentage of messages sent by user type that exchange inappropriate information (Y-axis) per max. ratio of inappropriate topics (X-axis)}
\label{fig:normsI}
\end{figure}

\begin{figure}[h!]
\begin{center}
\includegraphics[trim = 15mm 20mm 5mm 20mm,  clip,width=0.75\textwidth]{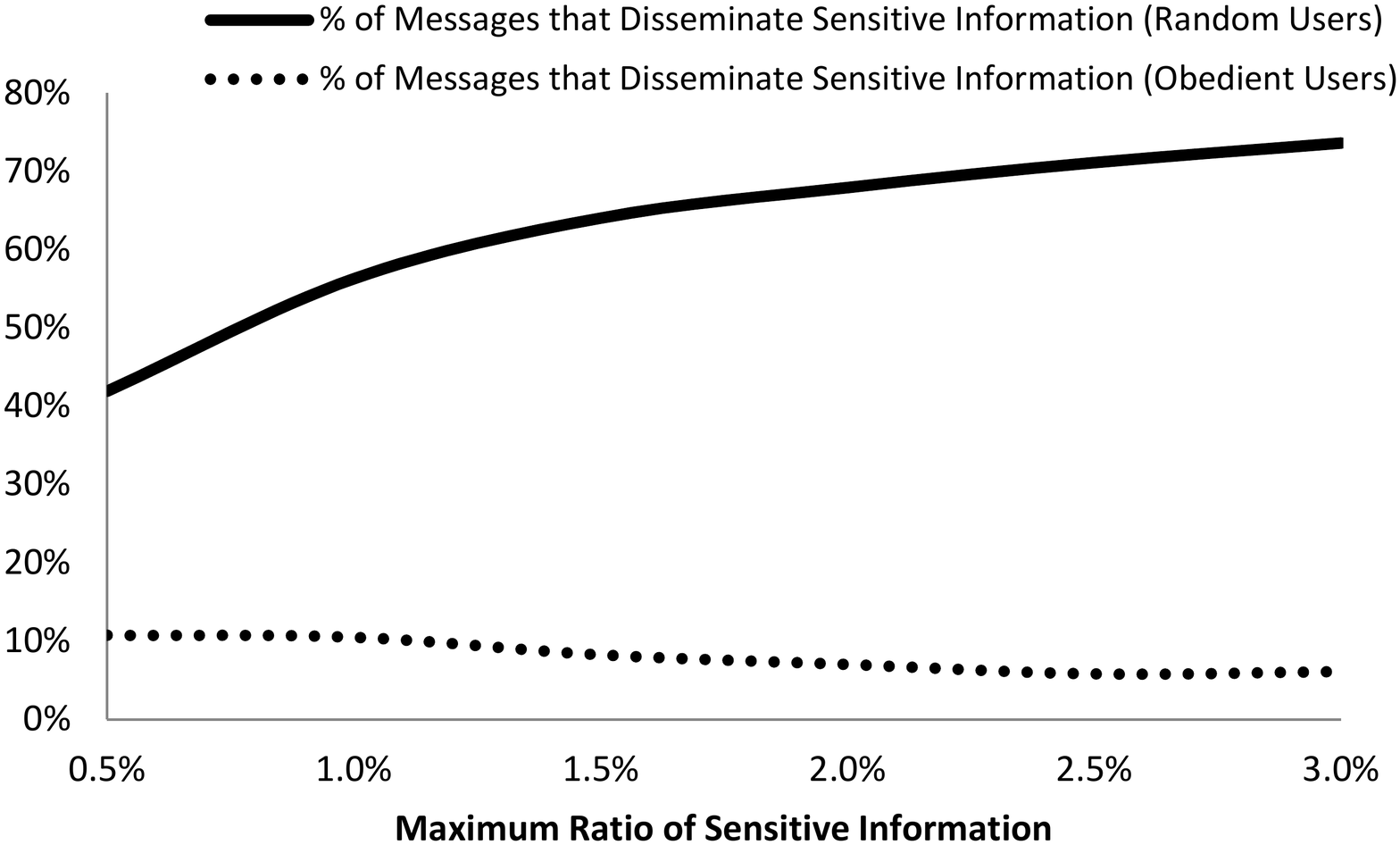} 
\end{center}
\vspace{-10pt}
\caption{Average percentage of messages sent by user type that disseminate sensitive information (Y-axis) per max. ratio of sensitive associations (X-axis)}
\label{fig:normsS}
\end{figure}

Figure \ref{fig:normsI} compares the behaviour exhibited by random users against obedient users in terms of the exchange of inappropriate information with respect to the maximum ratio of inappropriate topics. Figure \ref{fig:normsS} compares the behaviour exhibited by random users against obedient users in terms of the dissemination of sensitive information with respect to the maximum ratio of sensitive associations between users and topics. In case of random users, the average percentage of messages that violate some information sharing norm increases as the ratio of inappropriate topics or the ratio of sensitive associations increases. When the information to be discussed in a message is randomly selected, the probability of exchanging sensitive or inappropriate information increases as more information is considered illegal according to the information sharing norms. In case of obedient users, we obtained the opposite results (i.e., the average percentage of messages that violate some information sharing norm decreases slightly as the two ratios increase). Note that the amount of information (i.e., information sharing norms) that IAAs have to infer does not depend on the ratio of inappropriate topics or sensitive associations (i.e., for each topic IAAs have to infer to what degree it is sensitive or not, appropriate or not) and, as a result, the performance of our model is not affected negatively by the increase on the value of these ratios. On the contrary, IAAs obtain slightly better results when the ratio of inappropriate topics or the ratio of sensitive associations increases. This is explained by the fact that if more information is considered illegal according to information sharing norms, then compliant users (who always comply with information sharing norms) have less information to include in messages. Because of this, obedient users receive norm-complaint information at a higher frequency, which facilitates the inference of information sharing norms. Thus, IAAs' performance scales very well with increasing ratios of information sharing norms.

\subsubsection{Robustness against Malicious Users}
This experiment aimed to analyse the performance of IAAs in presence of malicious users; i.e., attackers who intend to inject fake messages so as to increase the appropriateness of a topic. Malicious users exchanged messages that included \textit{malicious} information (i.e., information relative to one inappropriate topic targeted by malicious users). In particular, we fixed the percentage of obedient users in each context to 10\%, and we varied the ratio of malicious users from 0\% to 90\%. The rest of users in each context were compliant users. We repeated this experiment with random users instead of obedient users as a baseline simulation. Each simulation was executed during 2000 steps and repeated 100 times. 

\begin{figure}[h!]
\begin{center}
\includegraphics[trim = 15mm 20mm 3mm 20mm,  clip,width=0.75\textwidth]{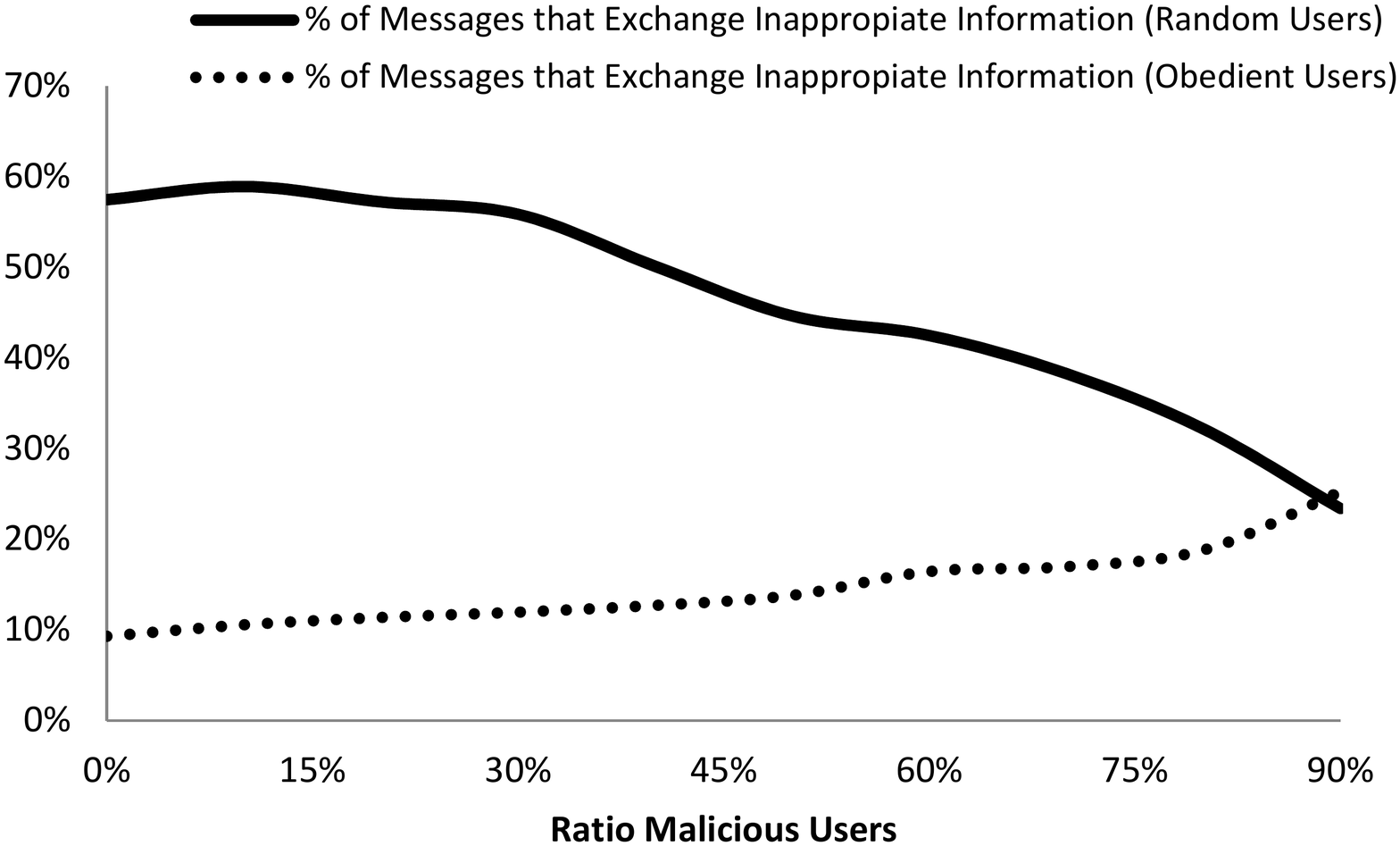} 
\end{center}
\vspace{-10pt}
\caption{Average percentage of messages sent by user type that exchange inappropriate information (Y-axis) per ratio malicious users (X-axis)}
\label{fig:malicious}
\end{figure}

Figure \ref{fig:malicious} shows the behaviour exhibited by random users and obedient users in terms of the exchange of inappropriate information. The average percentage of messages sent by random users that exchange inappropriate information decreases as the percentage of malicious users increases. This is explained by the fact that, malicious users always exchange the same malicious information, whereas compliant users exchange any information as long as this information is appropriate. However, a piece of information may be seen as appropriate in a given context and as inappropriate in another context. Thus, when compliant users are replaced by malicious users, less information is available in the network, so that random users have less information at their disposal, which makes them exchange less inappropriate information. On the contrary, in obedient users, the average percentage of messages that exchange inappropriate information increases as the percentage of malicious users increases. However, the average percentage of messages that exchange inappropriate information increases more slowly than malicious users do. In particular, IAAs prove to be robust against a single malicious user and small to medium-sized groups of malicious users. Indeed, obedient users behave as random users only when they are completely surrounded by malicious users. 

\subsubsection{Realistic User Populations}
The previous experiments were aimed at assessing the properties of IAAs with regards to some specific criteria. For this reason, we generated societies with extreme distributions of user types. However, these settings may not seem realistic; e.g., a situation in which all users are unconcerned about the norms is unlikely. In contrast, this section presents a more realistic simulation in which the distribution of the different user types match the results of empirical studies on privacy concerns among the general public \cite{taylor2003most,kumaraguru05}. In particular, and according to \cite{taylor2003most}, 
we have defined the ratio of obedient and compliant users (i.e., norm fundamentalists) as 26\% --- from which $15\%$ are compliant users to model the emergence of norms and the rest are obedient users; the ratio of relationship-based users (i.e., norm pragmatists) as 64\%; and the ratio of random users (i.e., norm unconcerned) as 10\%. 
Besides, recent studies demonstrate that in OSNs the majority of relationship ties are weak \cite{de2014facebook,zhao2010weak}. In particular, and according to \cite{zhao2010weak}, we have defined that 10\% of the friend relationships are close and trusted relationships. Finally, each simulation is executed during 4000 steps and has been repeated 100 times. 

\begin{figure}[h!]
\begin{center}
\vspace{-10pt}
\includegraphics[trim = 15mm 20mm 5mm 20mm,  clip,width=0.75\textwidth]{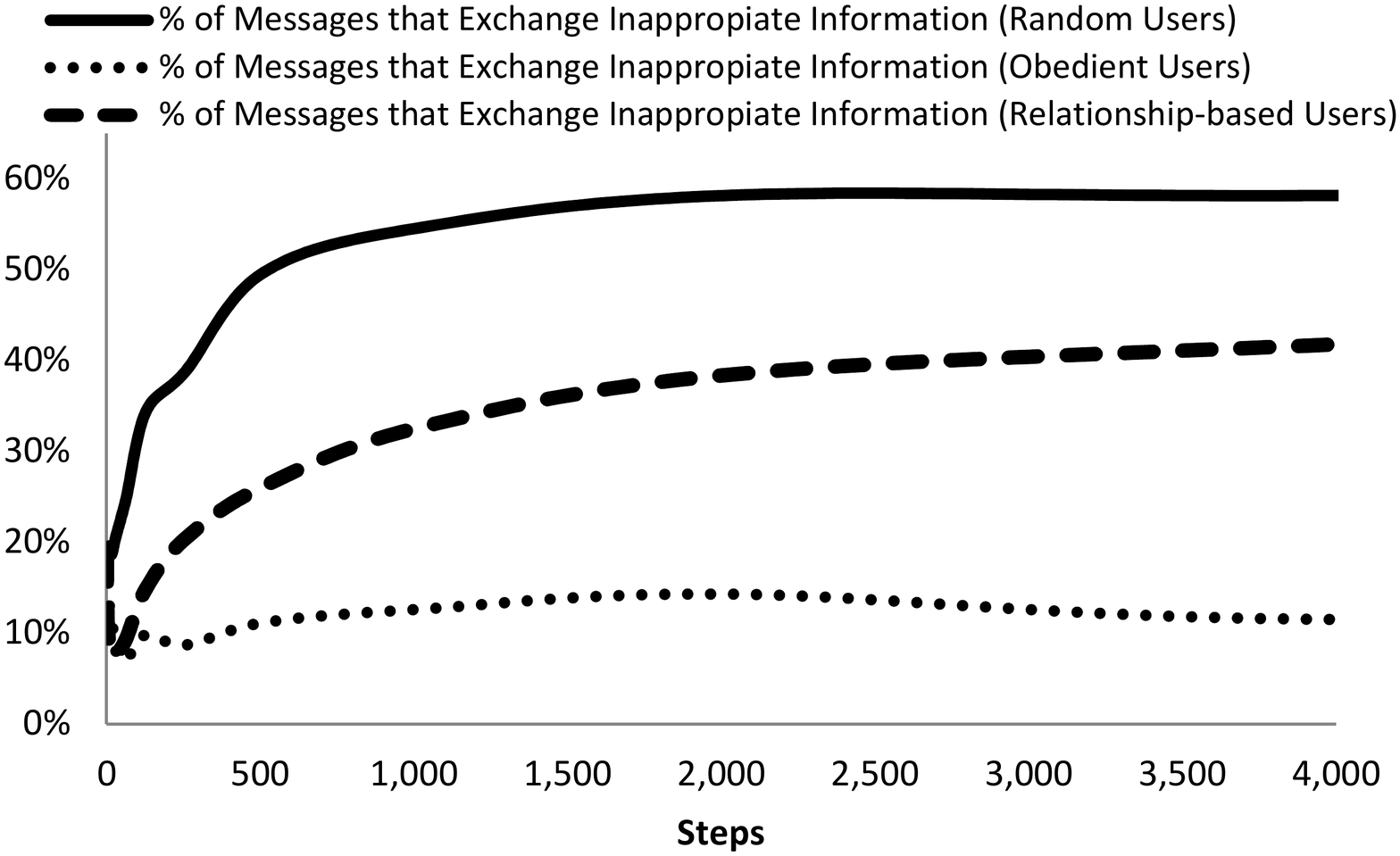} 
\end{center}
\vspace{-10pt}
\caption{Average percentage of messages sent by each user type that exchange inappropriate information (Y-axis) per number of steps (X-axis)}
\label{fig:exp7}
\end{figure}
\begin{figure}[h!]
\begin{center}
\vspace{-10pt}
\includegraphics[trim = 15mm 20mm 5mm 20mm,  clip,width=0.75\textwidth]{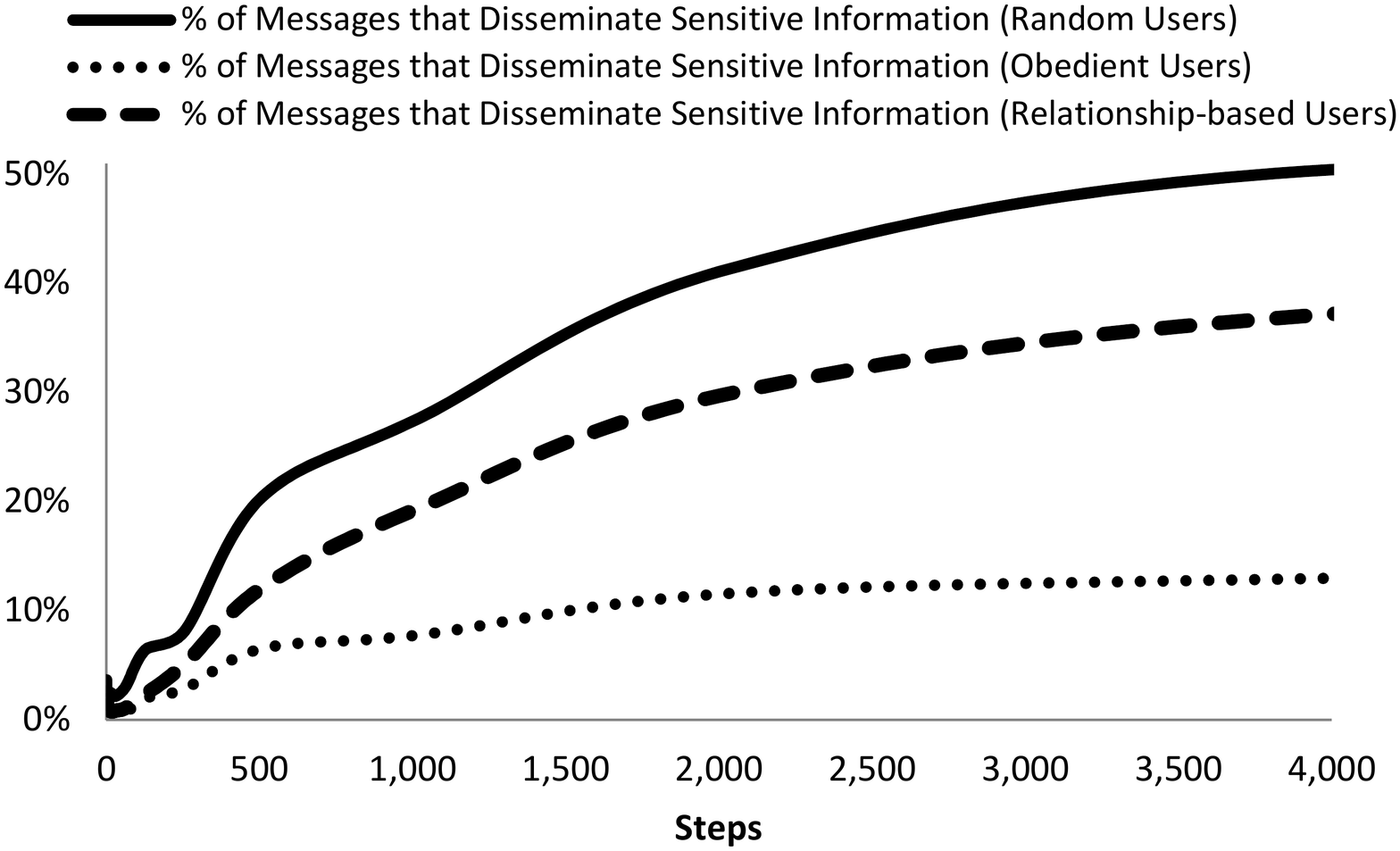} 
\end{center}
\vspace{-10pt}
\caption{Average percentage of messages sent by each user type that disseminate sensitive information (Y-axis) per number of steps (X-axis)}
\label{fig:exp8}
\end{figure}

Figure \ref{fig:exp7} displays the average percentage of messages sent by each type of users that exchange inappropriate information per number of steps that the simulation is executed. Similarly, Figure \ref{fig:exp8} displays the average percentage of messages sent by each type of users that disseminate sensitive information per number of steps that the simulation is executed. When the number of steps is very small, users have received few messages and they have a few pieces of information to be sent. Thus, the probability of violating information sharing norms is low. As the number of steps increases, users have received more pieces of information and they have more options to violate information sharing norms. As illustrated by these figures, all users that use IAAs (i.e., obedient and relationship-based) exchange significantly less inappropriate information and disseminate significantly less sensitive information than random users. Specifically, obedient users are the ones that violate less information sharing norms (obedient users achieve a reduction on norm violations over 70\%). This demonstrates that thanks to IAAs, obedient and relationship-based users, who don't know the information sharing norms, are able to comply with them. 

\section{Related Work}\label{sec:discussion}

\subsection{Contextual Integrity Modelling and Reasoning}
Previous work on computational models of contextual integrity proposed mechanisms for modelling and reasoning about contextual integrity principles. For example, Barth et al. \cite{barth2006privacy} formalized some aspects of contextual integrity assuming that there is a set of explicitly defined norms that determine what is permitted and forbidden, that the interactions take place in well-known contexts, and that interaction participants play a specific role in each interaction. In a more recent work, Krupa et al. \cite{krupa2012handling} proposed a framework to enforce information sharing norms in an electronic institution where norms, contexts and roles are explicitly defined. While these approaches seem appropriate for the kind of domains described in \cite{barth2006privacy} and \cite{krupa2012handling}, in OSNs there are not well-known contexts, there is not an explicit definition of the roles played by users and the exchange of information is governed by implicit information sharing norms\footnote{Note that these implicit information sharing norms are normality norms that define the behaviour that is consistent with the most common behaviour. In contrast, explicit information sharing norms define behaviour that is normative (i.e., moral).}.

\subsection{Access Control Models for OSNs}
The suitability of traditional access control models such as role-based access control \cite{sandhu1996role} for OSNs has been recently challenged on the basis that they cannot capture the inherent social nature of OSNs, such as social relationships and distance among users. To address this limitation, there are new paradigms that precisely emphasise the social aspects of OSNs. These new paradigms range from cryptographic privacy systems used to define and implement group access policies \cite{baden2009persona,guha2008noyb,lucas2008flybynight,tootoonchian2009lockr}, to relationship-based access control models \cite{fong2009privacy,fong2011relationship,fong2011preventing,bruns2012relationship,carminati2009enforcing,carminati2006rule,cheng2012user} that utilise a variety of features or aspects to characterise users' relationships and define access control decisions based on them. 

While these new models represent better frameworks than other traditional access control approaches to develop tools for defining and enforcing access control policies in OSNs, access control on its own is unlikely to be the complete and definitive solution for an appropriate privacy management in OSNs as users need awareness about access control decisions to fully understand the consequences of their access control policies \cite{kagal2010access,mondal2014,fogues2015open}. For instance, access control models are known to fail to prevent unintended disclosures \cite{bernstein2013quantifying}. This is where the implicit contextual integrity approach presented in this paper can make a difference by complementing these new models by learning implicit information sharing norms in OSNs and warning users where their defined access control policy may mean that inappropriate information would be exchanged or sensitive information would be disseminated across OSNs. 

\subsection{Disclosure Decision-Making Mechanisms}
In the related literature, the use of software endowed with disclosure decision-making mechanisms is not new. For example, several authors \cite{van2005value,krause2008utility} proposed mechanisms for computing the privacy-benefit trade-off of information disclosures in online interactions. The aim is to only disclose information when this trade-off renders appropriate results, i.e., where the utility of a particular disclosure is worth the privacy risks/consequences involved by performing the disclosure. However, these mechanisms have difficulties to deal with scenarios where the direct benefit of disclosing a piece of information is a priori unknown or difficult to express in economic terms, such as OSNs, in which disclosures are mostly driven by social factors \cite{houghton2010privacy}. In a more recent work, Such et al. \cite{Such201293} proposed a mechanism for entailing agents with capabilities to select the personal attributes of their users to be disclosed to other agents during interactions considering the increase on intimacy and privacy loss a disclosure may cause. However, this mechanism does not consider that the appropriateness of disclosures may vary from context to context, nor does it consider information disseminations. Also, this mechanism is fully automated whereas in our proposal IAAs only \emph{warn} users when information sharing norms may be violated but users have the last word on whether information is finally shared or not. 


\subsection{Topic Analysis}
Our proposal is based on the existence of a mechanism that is able to determine the topics discussed by the pieces of information that are exchanged in OSNs (i.e., the messages). Topics here can be categories  predefined by the underlying OSN infrastructure \cite{Peddinti:2014:CSU:2650286.2650765}; user-generated tags like Flickr categories \cite{nov2008drives}; or categories or tags extracted from images \cite{Sigurbjornsson:2008:FTR:1367497.1367542,wu2009distance}, videos \cite{Ballan:2011:ELS:2072298.2072060}, geolocation information \cite{lee2008tag}, or text \cite{rafeeque2011survey,thelwall2010sentiment,burnside2014one}.

A well-known problem of user-generated tags is that the tagging terms are often ambiguous, overly personalised and inexact \cite{guy2006tidying}. Indeed, current OSNs offer little or no synonym (i.e., different words, same meaning) or homonym (i.e., same word, different meanings) control \cite{golder2006,macgregor2006collaborative}. To overcome these problems, recent advances in the field of information sciences have proposed mechanisms for clustering semantically related tags (i.e., referred as topics in our paper) \cite{SPE:SPE2150,Radelaar:2014:ISE:2685119.2685126,giannakidou2008co}.

With regard to topic extraction from text, current research in the field of NLP has made advancements on analysing short messages present in OSN, microblogs, etc. For a review of these works see \cite{rafeeque2011survey}. Specifically, there are some NLP proposals endowed with new text-mining and analysis techniques that analyse short and informal messages with acceptable accuracy \cite{thelwall2010sentiment}. For example, the Relative Forensics toolkit\footnote{\url{http://www.relative-forensics.com}} employs, among other features, the 36 topics used in our experiments to analyse messages online messages. Specifically, the Relative Forensics toolkit has demonstrated to obtain a successful performance when detecting deceptive behaviour in online messages \cite{rashid}.

\subsection{Norm Learning}
Norm learning \cite{epstein2001learning} is the process of learning how to behave in a specific situation. In case of ONS information-sharing norms are implicit (i.e., there is not an explicit definition of what is sensitive or inappropriate), and supervised machine learning algorithms cannot be used to infer information sharing norms.

In the existing literature, \textit{social learning} \cite{DBLP:journals/aicom/CriadoAB11} of norms is defined as the process of inferring implicit social norms concurrently over repeated interactions with members of the social network. In most of the proposals on social learning, norms are inferred by analysing the outcomes of interactions and normative decisions in terms of utility \cite{sen2007emergence}. As previously mentioned, in OSN the benefit of exchanging information may be difficult to be determined in economic terms. In other proposals, norms are inferred by analysing explicit normative signals such as punishments, sanctions and rewards \cite{andrighetto}. These approaches cannot be used in OSN since implicit information sharing norms are product of informal social control that is rarely stated explicitly (e.g., sanctions) to unfriendly individuals. Other approaches \cite{epstein2001learning}, use imitation as a mechanism for learning social norms. In these proposals, the norms are inferred from the public behaviour exhibited by the majority of the members of the social network (or the majority of the members within an observation radius). A main drawback of imitation approaches is that all members are equally considered; i.e., they do not consider the existence of different social contexts with different social norms and the fact that users engage in relationships of different nature and strength.  

These unsupervised machine learning  approaches are unsuitable to be applied to ONS. This paper goes beyond these imitation approaches by considering the specific characteristics of OSNs, i.e., the existence of different and implicit social contexts with different information sharing norms, and the existence of different types of relationships between users. In particular, and to the best of our knowledge, our proposal is the first model of social learning of implicit information sharing social norms based on public and private information exchanges in OSNs.


\section{Conclusions}\label{sec:conclusion}

In this paper, we proposed the first computational model for implicit contextual integrity in OSNs. This model considers that in OSNs contexts are implicit, unknown a priori and they may change; user relationships evolve; and information sharing norms are implicit, i.e., they define the behaviour that is consistent with the most common behaviour. This model of implicit contextual integrity includes an information model for inferring the information sharing norms that are in force in OSNs and an Information Assistant Agent model to help users avoid exchanging inappropriate information and disseminating sensitive information across ONSs.

The experiments we conducted show that: (i) IAAs are able to infer information sharing norms even if a small proportion of the users follow the norms and in presence of malicious users; (ii) IAAs help to reduce the exchange of inappropriate information and the dissemination of sensitive information with only a partial view of the system and the information received and sent by their users; and (iii) IAAs minimise the burden to the users in terms of raising unnecessary alerts. 


As future work, we plan to equip IAAs with active mechanisms for detecting malicious users. As stated in the experiments, 
IAAs are robust against a single malicious user and small to medium-sized groups of malicious users. However, this robustness may not hold when an IAA is completely surrounded by a large number of malicious users. This does not mean that the user of the IAA would be more prone to disclose inappropriate information
, but that in the event she decides to disclose inappropriate information, the IAA may not alert her. Besides, note that these large groups of users would first need to be friends with IAA's user (or successfully infiltrate \emph{en masse} in her friend network), share contexts with her, and coordinate targeting a particular topic without the user noticing it. Automated approaches could be envisioned so that a single entity (whether individual or organisation) controls a large number of accounts to launch such attacks, but this can be mitigated using existing sybil defences \cite{alvisi2013sok,fong2011preventing}.  

\section*{References}
\bibliographystyle{abbrv}
\bibliography{references}

\begin{thebibliography}{10}

\bibitem{alvisi2013sok}
L.~Alvisi, A.~Clement, A.~Epasto, S.~Lattanzi, and A.~Panconesi.
\newblock Sok: The evolution of sybil defense via social networks.
\newblock In {\em Proceedings of the IEEE Symposium on Security and Privacy},
  pages 382--396, 2013.

\bibitem{baden2009persona}
R.~Baden, A.~Bender, N.~Spring, B.~Bhattacharjee, and D.~Starin.
\newblock Persona: an online social network with user-defined privacy.
\newblock In {\em ACM SIGCOMM Computer Communication Review}, volume~39, pages
  135--146. ACM, 2009.

\bibitem{Ballan:2011:ELS:2072298.2072060}
L.~Ballan, M.~Bertini, A.~Del~Bimbo, and G.~Serra.
\newblock Enriching and localizing semantic tags in internet videos.
\newblock In {\em Proceedings of the ACM International Conference on
  Multimedia}, pages 1541--1544, 2011.

\bibitem{barth2006privacy}
A.~Barth, A.~Datta, J.~C. Mitchell, and H.~Nissenbaum.
\newblock Privacy and contextual integrity: Framework and applications.
\newblock In {\em Proceedings of the IEEE Symposium on Security and Privacy},
  pages 184 -- 198, 2006.

\bibitem{bernstein2013quantifying}
M.~S. Bernstein, E.~Bakshy, M.~Burke, and B.~Karrer.
\newblock Quantifying the invisible audience in social networks.
\newblock In {\em Proceedings of the SIGCHI Conference on Human Factors in
  Computing Systems}, pages 21--30, 2013.

\bibitem{bilge2009all}
L.~Bilge, T.~Strufe, D.~Balzarotti, and E.~Kirda.
\newblock All your contacts are belong to us: automated identity theft attacks
  on social networks.
\newblock In {\em Proceedings of the International Conference on World Wide
  Web}, pages 551--560, 2009.

\bibitem{brooks2011modeling}
L.~Brooks, W.~Iba, and S.~Sen.
\newblock Modeling the emergence and convergence of norms.
\newblock In {\em IJCAI}, pages 97--102, 2011.

\bibitem{bruns2012relationship}
G.~Bruns, P.~W. Fong, I.~Siahaan, and M.~Huth.
\newblock Relationship-based access control: Its expression and enforcement
  through hybrid logic.
\newblock In {\em Proceedings of the ACM Conference on Data and Application
  Security and Privacy}, pages 117--124. ACM, 2012.

\bibitem{buccafurri2014driving}
F.~Buccafurri, G.~Lax, S.~Nicolazzo, A.~Nocera, and D.~Ursino.
\newblock Driving global team formation in social networks to obtain diversity.
\newblock In {\em Web Engineering}, pages 410--419. Springer, 2014.

\bibitem{buccafurri2013supporting}
F.~Buccafurri, G.~Lax, A.~Nocera, and D.~Ursino.
\newblock Supporting information spread in a social internetworking scenario.
\newblock In {\em New Frontiers in Mining Complex Patterns}, pages 200--214.
  Springer, 2013.

\bibitem{burke2014growing}
M.~Burke and R.~E. Kraut.
\newblock Growing closer on facebook: changes in tie strength through social
  network site use.
\newblock In {\em Proceedings of the 32nd annual ACM conference on Human
  factors in computing systems}, pages 4187--4196, 2014.

\bibitem{burnside2014one}
G.~Burnside, D.~Milioris, P.~Jacquet, et~al.
\newblock One day in twitter: Topic detection via joint complexity.
\newblock In {\em SNOW Data Challenge}, pages 41--48, 2014.

\bibitem{carminati2006rule}
B.~Carminati, E.~Ferrari, and A.~Perego.
\newblock Rule-based access control for social networks.
\newblock In R.~Meersman, Z.~Tari, and P.~Herrero, editors, {\em On the Move to
  Meaningful Internet Systems 2006: OTM 2006 Workshops}, volume 4278 of {\em
  Lecture Notes in Computer Science}, pages 1734--1744. Springer Berlin
  Heidelberg, 2006.

\bibitem{carminati2009enforcing}
B.~Carminati, E.~Ferrari, and A.~Perego.
\newblock Enforcing access control in web-based social networks.
\newblock {\em ACM Transactions on Information and System Security},
  13(1):6:1--6:38, Nov. 2009.

\bibitem{DBLP:conf/socialcom/ChengRMK11}
J.~Cheng, D.~M. Romero, B.~Meeder, and J.~M. Kleinberg.
\newblock Predicting reciprocity in social networks.
\newblock In {\em Privacy, Security, Risk and Trust, IEEE International
  Conference on Social Computing}, pages 49--56, 2011.

\bibitem{cheng2012user}
Y.~Cheng, J.~Park, and R.~Sandhu.
\newblock A user-to-user relationship-based access control model for online
  social networks.
\newblock In N.~Cuppens-Boulahia, F.~Cuppens, and J.~Garcia-Alfaro, editors,
  {\em Data and Applications Security and Privacy XXVI}, volume 7371 of {\em
  Lecture Notes in Computer Science}, pages 8--24. Springer Berlin Heidelberg,
  2012.

\bibitem{DBLP:journals/aicom/CriadoAB11}
N.~Criado, E.~Argente, and V.~J. Botti.
\newblock Open issues for normative multi-agent systems.
\newblock {\em {AI} Communications}, 24(3):233--264, 2011.

\bibitem{hargittai2010facebook}
B.~Danah and E.~Hargittai.
\newblock Facebook privacy settings: Who cares?
\newblock {\em First Monday}, 15(8), 2010.

\bibitem{danezis2009inferring}
G.~Danezis.
\newblock Inferring privacy policies for social networking services.
\newblock In {\em Proceedings of the ACM workshop on Security and artificial
  intelligence}, pages 5--10, 2009.

\bibitem{dasgupta1997evolutionary}
D.~Dasgupta and Z.~Michalewicz.
\newblock {\em Evolutionary algorithms in engineering applications}.
\newblock Springer Science \& Business Media, 1997.

\bibitem{de2014facebook}
P.~De~Meo, E.~Ferrara, G.~Fiumara, and A.~Provetti.
\newblock On facebook, most ties are weak.
\newblock {\em Communications of the ACM}, 57(11):78--84, 2014.

\bibitem{elster1989social}
J.~Elster.
\newblock Social norms and economic theory.
\newblock {\em The Journal of Economic Perspectives}, pages 99--117, 1989.

\bibitem{epstein2001learning}
J.~M. Epstein.
\newblock Learning to be thoughtless: Social norms and individual computation.
\newblock {\em Computational Economics}, 18(1):9--24, 2001.

\bibitem{SPE:SPE2150}
D.~Eynard, L.~Mazzola, and A.~Dattolo.
\newblock Exploiting tag similarities to discover synonyms and
  homonyms in folksonomies.
\newblock {\em Software: Practice and Experience}, 43(12):1437--1457, 2013.

\bibitem{fogues2015open}
R.~Fogues, J.~M. Such, A.~Espinosa, and A.~Garcia-Fornes.
\newblock Open challenges in relationship-based privacy mechanisms for social
  network services.
\newblock {\em International Journal of Human-Computer Interaction},
  (just-accepted), 2015.

\bibitem{fogues2014bff}
R.~L. Fogues, J.~M. Such, A.~Espinosa, and A.~Garcia-Fornes.
\newblock Bff: A tool for eliciting tie strength and user communities in social
  networking services.
\newblock {\em Information Systems Frontiers}, 16(2):225--237, 2014.

\bibitem{fong2009privacy}
P.~Fong, M.~Anwar, and Z.~Zhao.
\newblock A privacy preservation model for facebook-style social network
  systems.
\newblock In M.~Backes and P.~Ning, editors, {\em Computer Security – ESORICS
  2009}, volume 5789 of {\em Lecture Notes in Computer Science}, pages
  303--320. Springer Berlin Heidelberg, 2009.

\bibitem{fong2011preventing}
P.~W. Fong.
\newblock Preventing sybil attacks by privilege attenuation: A design principle
  for social network systems.
\newblock In {\em IEEE Symposium on Security and Privacy}, pages 263--278,
  2011.

\bibitem{fong2011relationship}
P.~W. Fong.
\newblock Relationship-based access control: Protection model and policy
  language.
\newblock In {\em Proceedings of the ACM Conference on Data and Application
  Security and Privacy}, pages 191--202, 2011.

\bibitem{fowler}
G.~A. Fowler.
\newblock When the most personal secrets get outed on facebook.
\newblock \url{http://online.wsj.com/articles/
  SB10000872396390444165804578008740578200224}, Accessed: Nov, 2014.

\bibitem{giannakidou2008co}
E.~Giannakidou, V.~Koutsonikola, A.~Vakali, and I.~Kompatsiaris.
\newblock Co-clustering tags and social data sources.
\newblock In {\em Web-Age Information Management, 2008. WAIM'08. The Ninth
  International Conference on}, pages 317--324. IEEE, 2008.

\bibitem{girvan2002community}
M.~Girvan and M.~E. Newman.
\newblock Community structure in social and biological networks.
\newblock {\em Proceedings of the National Academy of Sciences},
  99(12):7821--7826, 2002.

\bibitem{golder2006}
S.~Golder and B.~A. Huberman.
\newblock Usage patterns of collaborative tagging systems.
\newblock {\em Journal of Information Science}, 32(2):198--208, 2006.

\bibitem{greene2006self}
K.~Greene, V.~J. Derlega, and A.~Mathews.
\newblock Self-disclosure in personal relationships.
\newblock {\em The Cambridge handbook of personal relationships}, pages
  409--427, 2006.

\bibitem{gross2005information}
R.~Gross and A.~Acquisti.
\newblock Information revelation and privacy in online social networks.
\newblock In {\em Proceedings of the ACM workshop on Privacy in the electronic
  society}, pages 71--80, 2005.

\bibitem{guha2008noyb}
S.~Guha, K.~Tang, and P.~Francis.
\newblock Noyb: Privacy in online social networks.
\newblock In {\em Proceedings of the first workshop on Online social networks},
  pages 49--54. ACM, 2008.

\bibitem{guy2006tidying}
M.~Guy and E.~Tonkin.
\newblock Tidying up tags.
\newblock {\em D-lib Magazine}, 12(1):1082--9873, 2006.

\bibitem{houghton2010privacy}
D.~J. Houghton and A.~N. Joinson.
\newblock Privacy, social network sites, and social relations.
\newblock {\em Journal of Technology in Human Services}, 28(1-2):74--94, 2010.

\bibitem{iba2009applied}
H.~Iba, Y.~Hasegawa, and T.~K. Paul.
\newblock {\em Applied genetic programming and machine learning}.
\newblock cRc Press, 2009.

\bibitem{jagatic2007social}
T.~N. Jagatic, N.~A. Johnson, M.~Jakobsson, and F.~Menczer.
\newblock Social phishing.
\newblock {\em Communications of the ACM}, 50(10):94--100, 2007.

\bibitem{kagal2010access}
L.~Kagal and H.~Abelson.
\newblock Access control is an inadequate framework for privacy protection.
\newblock In {\em W3C Privacy Workshop}, pages 1--6, 2010.

\bibitem{krause2008utility}
A.~Krause and E.~Horvitz.
\newblock A utility-theoretic approach to privacy and personalization.
\newblock In {\em Proceedings of the AAAI Conference on Artificial
  Intelligence}, volume~8, pages 1181--1188, 2008.

\bibitem{krupa2012handling}
Y.~Krupa and L.~Vercouter.
\newblock Handling privacy as contextuxal integrity in decentralized virtual
  communities: The privacias framework.
\newblock {\em Web Intelligence and Agent Systems}, 10(1):105--116, 2012.

\bibitem{kumaraguru05}
P.~Kumaraguru and L.~Cranor.
\newblock Privacy indexes: A survey of westin's studies.
\newblock Technical Report CMU-ISRI-5-138, Carnegie Mellon University, School
  of Computer Science, Institute for Software Research International, 2005.

\bibitem{lee2008tag}
S.~S. Lee, D.~Won, and D.~McLeod.
\newblock Tag-geotag correlation in social networks.
\newblock In {\em Proceedings of the ACM Workshop on Search in Social Media},
  pages 59--66, 2008.

\bibitem{lucas2008flybynight}
M.~M. Lucas and N.~Borisov.
\newblock Flybynight: mitigating the privacy risks of social networking.
\newblock In {\em Proceedings of the 7th ACM workshop on Privacy in the
  electronic society}, pages 1--8. ACM, 2008.

\bibitem{lyndon2011college}
A.~Lyndon, J.~Bonds-Raacke, and A.~D. Cratty.
\newblock College students' facebook stalking of ex-partners.
\newblock {\em Cyberpsychology, Behavior, and Social Networking},
  14(12):711--716, 2011.

\bibitem{macgregor2006collaborative}
G.~Macgregor and E.~McCulloch.
\newblock Collaborative tagging as a knowledge organisation and resource
  discovery tool.
\newblock {\em Library review}, 55(5):291--300, 2006.

\bibitem{mislove2007measurement}
A.~Mislove, M.~Marcon, K.~P. Gummadi, P.~Druschel, and B.~Bhattacharjee.
\newblock Measurement and analysis of online social networks.
\newblock In {\em Proceedings of the ACM SIGCOMM conference on Internet
  measurement}, pages 29--42, 2007.

\bibitem{mondal2014}
M.~Mondal, P.~Druschel, K.~P. Gummadi, and A.~Mislove.
\newblock {Beyond Access Control: Managing Online Privacy via Exposure}.
\newblock In {\em {Proceedings of the Workshop on Useable Security}}, pages
  1--6, {2014}.

\bibitem{nakano2005self}
T.~Nakano and T.~Suda.
\newblock Self-organizing network services with evolutionary adaptation.
\newblock {\em Neural Networks, IEEE Transactions on}, 16(5):1269--1278, 2005.

\bibitem{nissenbaum2004privacy}
H.~Nissenbaum.
\newblock Privacy as contextual integrity.
\newblock {\em Washington Law Review}, 79(1):119--158, 2004.

\bibitem{nov2008drives}
O.~Nov, M.~Naaman, and C.~Ye.
\newblock What drives content tagging: the case of photos on flickr.
\newblock In {\em Proceedings of the SIGCHI conference on Human factors in
  computing systems}, pages 1097--1100, 2008.

\bibitem{Peddinti:2014:CSU:2650286.2650765}
S.~T. Peddinti, A.~Korolova, E.~Bursztein, and G.~Sampemane.
\newblock Cloak and swagger: Understanding data sensitivity through the lens of
  user anonymity.
\newblock In {\em Proceedings of the IEEE Symposium on Security and Privacy},
  pages 493--508, 2014.

\bibitem{pike2011fired}
G.~H. Pike.
\newblock Fired over facebook.
\newblock {\em Information Today}, 28(4):26--26, 2011.

\bibitem{Radelaar:2014:ISE:2685119.2685126}
J.~Radelaar, A.-J. Boor, D.~Vandic, J.-W. Van~Dam, and F.~Fasincar.
\newblock Improving search and exploration in tag spaces using automated tag
  clustering.
\newblock {\em Journal of Web Engineering}, 13(3-4):277--301, July 2014.

\bibitem{rafeeque2011survey}
P.~Rafeeque and S.~Sendhilkumar.
\newblock A survey on short text analysis in web.
\newblock In {\em Proceedings of the International Conference on Advanced
  Computing}, pages 365--371, 2011.

\bibitem{rashid}
A.~Rashid, A.~Baron, P.~Rayson, C.~May-Chahal, P.~Greenwood, and J.~Walkerdine.
\newblock Who am i? analyzing digital personas in cybercrime investigations.
\newblock {\em IEEE Computer}, 46(4):54--61, 2013.

\bibitem{raynes2006hyperfriends}
K.~Raynes-Goldie and D.~Fono.
\newblock Hyperfriends and beyond: Friendship and social norms on livejournal.
\newblock {\em Internet Research Annual}, 4:8, 2006.

\bibitem{Rayson}
P.~Rayson.
\newblock {From key words to key semantic domains}.
\newblock {\em International Journal of Corpus Linguistics}, 13(4):519--549,
  2008.

\bibitem{ross2002fuzzy}
T.~J. Ross, J.~M. Booker, and W.~J. Parkinson.
\newblock {\em Fuzzy logic and probability applications: bridging the gap},
  volume~11.
\newblock SIAM, 2002.

\bibitem{rosvall2007information}
M.~Rosvall and C.~T. Bergstrom.
\newblock An information-theoretic framework for resolving community structure
  in complex networks.
\newblock {\em Proceedingsof the National Academy of Sciences},
  104(18):7327--7331, 2007.

\bibitem{ruedy2007repercussions}
M.~C. Ruedy.
\newblock Repercussions of a myspace teen suicide: Should anti-cyberbullying
  laws be created.
\newblock {\em North Carolina Journal of Law \& Technology}, 9:323--346, 2007.

\bibitem{sandhu1996role}
R.~S. Sandhu, E.~J. Coyne, H.~L. Feinstein, and C.~E. Youman.
\newblock Role-based access control models.
\newblock {\em Computer}, 29(2):38--47, 1996.

\bibitem{sen2007emergence}
S.~Sen and S.~Airiau.
\newblock Emergence of norms through social learning.
\newblock In {\em Proceedings of the International Joint Conference on
  Artifical Intelligence}, pages 1507--1512, 2007.

\bibitem{Sigurbjornsson:2008:FTR:1367497.1367542}
B.~Sigurbj\"{o}rnsson and R.~van Zwol.
\newblock Flickr tag recommendation based on collective knowledge.
\newblock In {\em Proceedings of the International Conference on World Wide
  Web}, pages 327--336, 2008.

\bibitem{stevens}
J.~Stevens.
\newblock The facebook divorces: Social network site is cited in 'a third of
  splits'.
\newblock
  \url{http://www.dailymail.co.uk/femail/article-2080398/Facebook-cited-THIRD-divorces.html},
  Accessed: Nov, 2014.

\bibitem{strahilevitz2005social}
L.~J. Strahilevitz.
\newblock A social networks theory of privacy.
\newblock In {\em American Law \& Economics Association Annual Meetings}, pages
  919--988, 2005.

\bibitem{strater2008strategies}
K.~Strater and H.~R. Lipford.
\newblock Strategies and struggles with privacy in an online social networking
  community.
\newblock In {\em Proceedings of the HCI Group Annual Conference on People and
  Computers: Culture, Creativity, Interaction}, pages 111--119, 2008.

\bibitem{stutzman2013silent}
F.~Stutzman, R.~Gross, and A.~Acquisti.
\newblock Silent listeners: The evolution of privacy and disclosure on
  facebook.
\newblock {\em Journal of Privacy and Confidentiality}, 4(2), 2013.

\bibitem{Such201293}
J.~M. Such, A.~Espinosa, A.~Garcia-Fornes, and C.~Sierra.
\newblock Self-disclosure decision making based on intimacy and privacy.
\newblock {\em Information Sciences}, 211(0):93 -- 111, 2012.

\bibitem{taylor2003most}
H.~Taylor.
\newblock Most people are ‘privacy pragmatists’ who, while concerned about
  privacy, will sometimes trade it off for other benefits.
\newblock {\em The Harris Poll}, 17(19):44, 2003.

\bibitem{dma2012}
{The {D}irect {M}arketing {A}ssociation DMA (UK) Ltd}.
\newblock {\em Data privacy: What the consumer really thinks}.
\newblock 2012.

\bibitem{thelwall2010sentiment}
M.~Thelwall, K.~Buckley, G.~Paltoglou, D.~Cai, and A.~Kappas.
\newblock Sentiment strength detection in short informal text.
\newblock {\em Journal of the American Society for Information Science and
  Technology}, 61(12):2544--2558, 2010.

\bibitem{tootoonchian2009lockr}
A.~Tootoonchian, S.~Saroiu, Y.~Ganjali, and A.~Wolman.
\newblock Lockr: better privacy for social networks.
\newblock In {\em Proceedings of the 5th international conference on Emerging
  networking experiments and technologies}, pages 169--180. ACM, 2009.

\bibitem{van2005value}
S.~van Otterloo.
\newblock The value of privacy: optimal strategies for privacy minded agents.
\newblock In {\em Proceedings of the International Conference on Autonomous
  Agents and Multiagent Systems}, pages 1015--1022, 2005.

\bibitem{andrighetto}
D.~Villatoro, G.~Andrighetto, J.~Sabater{-}Mir, and R.~Conte.
\newblock Dynamic sanctioning for robust and cost-efficient norm compliance.
\newblock In {\em Proceedings of the International Joint Conference on
  Artificial Intelligence}, pages 414--419, 2011.

\bibitem{vorvoreanu2009perceptions}
M.~Vorvoreanu.
\newblock Perceptions of corporations on facebook: An analysis of facebook
  social norms.
\newblock {\em Journal of New Communications Research}, 4(1):67--86, 2009.

\bibitem{wang2011regretted}
Y.~Wang, G.~Norcie, S.~Komanduri, A.~Acquisti, P.~G. Leon, and L.~F. Cranor.
\newblock I regretted the minute i pressed share: A qualitative study of
  regrets on facebook.
\newblock In {\em Proceedings of the Seventh Symposium on Usable Privacy and
  Security}, page~10, 2011.

\bibitem{wu2009distance}
L.~Wu, S.~C. Hoi, R.~Jin, J.~Zhu, and N.~Yu.
\newblock Distance metric learning from uncertain side information with
  application to automated photo tagging.
\newblock In {\em Proceedings of the ACM International Conference on
  Multimedia}, pages 135--144, 2009.

\bibitem{zhao2010weak}
J.~Zhao, J.~Wu, and K.~Xu.
\newblock Weak ties: Subtle role of information diffusion in online social
  networks.
\newblock {\em Physical Review E}, 82(1):1--8, 2010.

\end{thebibliography}

\end{document}